\documentclass[fleqn,usenatbib]{mnras}

\usepackage{newtxtext,newtxmath}

\usepackage[T1]{fontenc}

\DeclareRobustCommand{\VAN}[3]{#2}
\let\VANthebibliography\thebibliography
\def\thebibliography{\DeclareRobustCommand{\VAN}[3]{##3}\VANthebibliography}

\usepackage{graphicx}	
\usepackage{amsmath}	

\newcommand{\psrzero}{PSR~J0058$-$7218}
\newcommand{\psrfive}{PSR~J0537$-$6910}
\newcommand{\psrzerofive}{PSR~B0540$-$69}
\newcommand{\psrfour}{PSR~J1412+7922}
\newcommand{\psreightone}{PSR~J1811$-$1925}
\newcommand{\psreightthree}{PSR~J1813$-$1749}
\newcommand{\psreight}{PSR~J1849$-$0001}
\newcommand{\Pdot}{\dot{P}}
\newcommand{\nudot}{\dot{\nu}}
\newcommand{\nuddot}{\ddot{\nu}}
\newcommand{\nudddot}{\dddot{\nu}}
\newcommand{\nig}{n_{\rm ig}}
\newcommand{\tauc}{\tau_{\rm c}}
\newcommand{\Edot}{\dot{E}}
\newcommand{\NH}{N_{\rm H}}
\newcommand{\tg}{t_{\rm g}}

\title[Timing seven young energetic X-ray pulsars]{New and updated timing models for seven young energetic X-ray pulsars, including the Big Glitcher \psrfive}

\author[W. C. G. Ho et al.]{
Wynn C. G. Ho,$^{1}$\thanks{E-mail: who@haverford.edu}
Lucien Kuiper,$^{2}$
Crist\'obal M. Espinoza,$^{3,4}$
Timothy Leon,$^{1}$
Bennett Waybright,$^{1}$
\newauthor
Sebastien Guillot,$^{5,6}$
Zaven Arzoumanian,$^{7}$
Slavko Bogdanov,$^{8}$
and
Alice K. Harding$^{9}$
\\
$^{1}$Department of Physics and Astronomy, Haverford College, 370 Lancaster Avenue, Haverford, PA 19041, USA\\
$^{2}$Space Research Organisation Netherlands, Niels Bohrweg 4, 2333 CA, Leiden, Netherlands\\
$^{3}$Departamento de F\'isica, Universidad de Santiago de Chile (USACH), Av. V\'ictor Jara 3493, Estaci\'on Central, Chile\\
$^{4}$Center for Interdisciplinary Research in Astrophysics and Space Sciences (CIRAS), Universidad de Santiago de Chile, Santiago, Chile\\
$^{5}$IRAP, CNRS, 9 avenue du Colonel Roche, BP 44346, 31028 Toulouse Cedex 4, France\\
$^{6}$Universit\'e de Toulouse, CNES, UPS-OMP, 31028 Toulouse, France\\
$^{7}$X-Ray Astrophysics Laboratory, NASA Goddard Space Flight Center, Greenbelt, MD, 20771, USA\\
$^{8}$Columbia Astrophysics Laboratory, Columbia University, 550 West 120th Street, New York, NY 10027, USA\\
$^{9}$Theoretical Division, Los Alamos National Laboratory, Los Alamos, NM 87545, USA\\
}

\date{Accepted 2025 December 19. Received 2025 December 4; in original form 2025 October 28}

\pubyear{2026}

\begin{document}
\label{firstpage}
\pagerange{\pageref{firstpage}--\pageref{lastpage}}
\maketitle

\begin{abstract}
We present new timing models and update our previous ones for the
rotational evolution of seven young energetic pulsars, including four
of the top five in spin-down luminosity $\Edot$ among all known pulsars.
For each of the six pulsars that were monitored on a regular basis by NICER,
their rotation phase-connected timing model covers the entire period of
NICER observations, in many cases from 2017--2025.
For \psrzero, which was only identified in 2021, we extend the baseline of its
timing model by 3 years and report detections of its first three glitches.
The timing model for \psrfive\ over the entire 8 years of
NICER monitoring is presented, including a total of 23 glitches;
we also report its spin frequency and pulsed spectrum from a 2016 NuSTAR
observation.
For \psrzerofive, its complete timing model from 2015--2025 is provided,
including a braking index evolution from near 0 to 1.6 during this period.
The 8-year timing model for \psrfour\ (also known as Calvera) is reported,
which includes a position that is consistent with that measured from imaging.
For \psreightone, we present its 3.5-year timing model.
For \psreightthree, its incoherent timing model is extended through early
2025 using new Chandra observations.
For \psreight, its 7-year timing model is provided, including
a position that is consistent with and more accurate than its imaging position
and its first glitch that is one of the largest ever measured.
Our timing models of these seven X-ray pulsars enable their study at
other energies and in gravitational wave data.
\end{abstract}




\section{Introduction} \label{sec:intro}

Long-term monitoring and timing observations to measure the rotational
evolution of pulsars are crucial for studying their individual properties,
for classifying them, and for using them to study fundamental physics
and gravity.
For example, sudden measured changes in the spin frequency,
also known as glitches, of some pulsars provide crucial insights into
neutron star crust properties and superfluidity in dense nuclear matter
(see, e.g., \citealt{antonopoulouetal22}, for review).
Precision timing of millisecond pulsars allowed detection of low frequency
gravitational waves produced by binary inspirals of supermassive
black holes \citep{agazieetal23}.
Pulsar timing also enables the most sensitive searches for high frequency
gravitational waves produced by isolated and binary pulsars
(e.g., \citealt{abacetal25}).

\begin{table*}
\centering
\caption{Properties of pulsars considered in present work.
Spin period $P$ and spin period time derivative $\Pdot$,
spin-down luminosity $\Edot=4.0\times10^{46}\mbox{ erg s$^{-1}$ }\Pdot/P^3$,
magnetic field $B=3.2\times10^{19}\mbox{ G }(P\Pdot)^{1/2}$,
characteristic age $\tauc\equiv P/2\Pdot$,
supernova remnant (SNR) association and age, and distance $d$.
References: [1] \citet{owenetal11}, [2]: \citet{graczyketal20},
[3]: \citet{chenetal06}, [4]: \citet{pietrzynskietal19},
[5]: \citet{ariasetal22}, [6]: \citet{grecoetal25}, [7]: \citet{mereghettietal21},
[8]: \citet{borkowskietal16}, [9] \citet{kilpatricketal16},
[10]: \citet{broganetal05}, [11]: \citet{dzibrodriguez21}, [12]: \citet{camiloetal21},
[13]: \citet{gotthelfetal11}.
}
\label{tab:psr}
\begin{tabular}{lcccccccc}
  \hline
Pulsar & $P$ & $\Pdot$ & $\Edot$ & $B$ & $\tauc$ & SNR & SNR age & $d$ \\
& (ms) & ($10^{-14}\mbox{ s s$^{-1}$}$) & ($10^{37}\mbox{ erg s$^{-1}$}$) & ($10^{12}\mbox{ G}$) & (kyr) & & (kyr) & (kpc) \\
  \hline
\psrzero & 21.8 & 2.95 & 11 & 0.81 & 11.7 & IKT~16 & 14.7 [1] & 62 [2] \\
\psrfive & 16.2 & 5.21 & 49 & 0.93 & 4.91 & N157B & 1--5 [3] & 49.6 [4] \\
\psrzerofive & 50.7 & 47.9 & 15 & 5.0 & 1.68 & 0540$-$69.3 & & 49.6 [4] \\
\psrfour & 59.2 & 0.330 & 0.064 & 0.45 & 285 & G118.4+37.0 & 10--20 [5,6] & 3--5 [6,7] \\
\psreightone & 64.7 & 4.40 & 0.65 & 1.7 & 24 & G11.2$-$0.3 & 1.4--2.4 [8] & 7.2 [9] \\
\psreightthree & 44.7 & 12.7 & 5.7 & 2.4 & 5.58 & G12.82$-$0.02 & 1--2.2 [10,11] & 6--14 [12] \\
\psreight & 38.5 & 1.42 & 0.99 & 0.75 & 43.1 & & & 7 [13] \\
  \hline
\end{tabular}
\end{table*}

\begin{figure}
\includegraphics[width=\columnwidth]{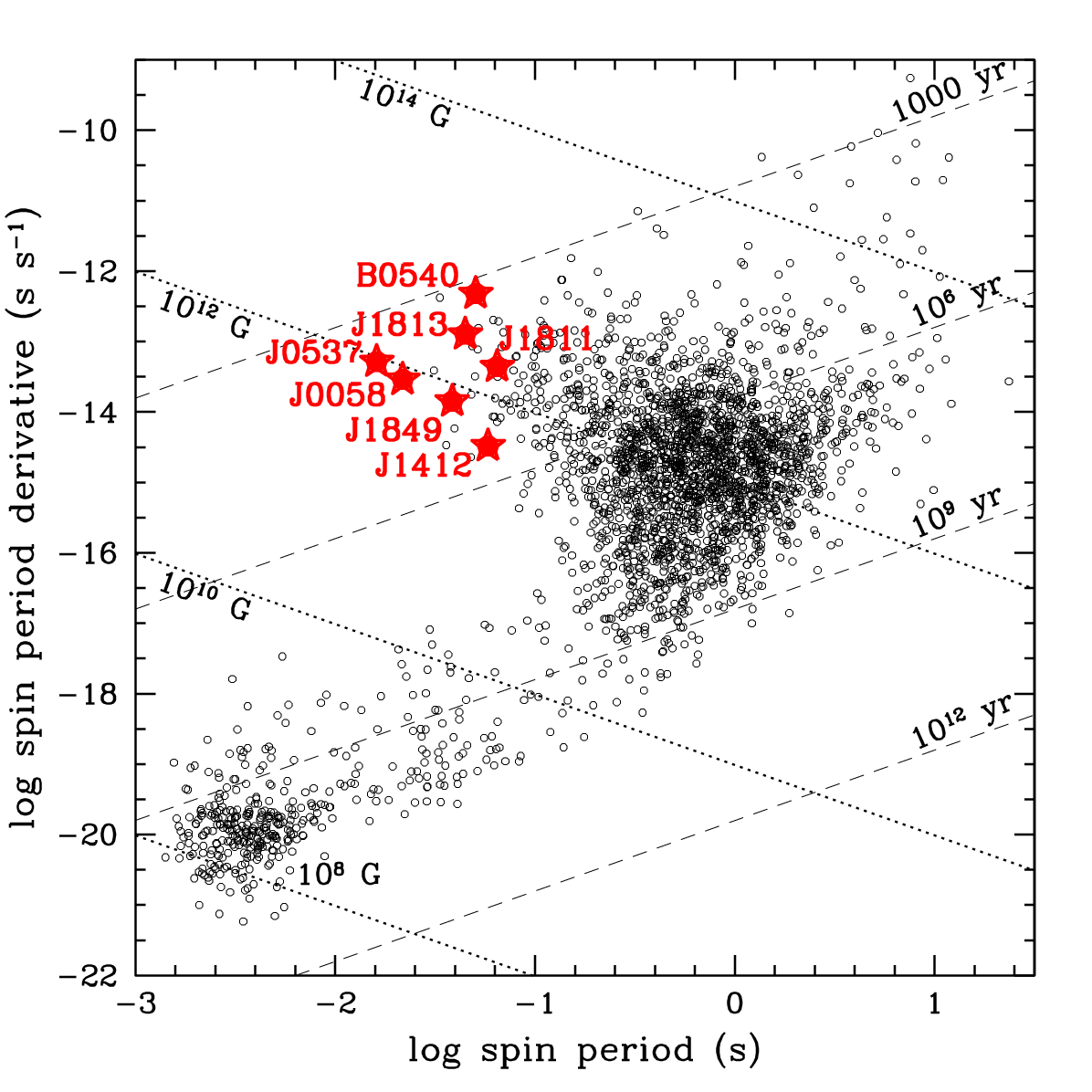}
\caption{
Pulsar spin period $P$ and spin period time derivative $\Pdot$.
Circles denote pulsars whose values are taken from the ATNF Pulsar Catalogue
(\citealt{manchesteretal05}, version~2.3.0), and stars indicate pulsars
considered in this work (see Table~\ref{tab:psr}).
Dashed lines indicate characteristic age $\tauc\equiv P/2\Pdot$, and
dotted lines indicate magnetic field strength
$B=3.2\times10^{19}\mbox{ G }(P\Pdot)^{1/2}$.}
\label{fig:ppdot}
\end{figure}

Here we present updates of the timing models, or ephemerides, of six
young energetic pulsars that were reported in \citet{hoetal22} and
\citet{espinozaetal24}, as well as the timing model of one other
young energetic pulsar, \psreightone.
For six of the seven timing models, the models are rotation
phase-connected and are therefore tracking each individual rotation
of the pulsar.
The seven pulsars are predominantly only observable in the X-ray, and the
data used here are primarily from regular monitoring observations made by the
Neutron Star Interior Composition Explorer (NICER; \citealt{gendreauetal16})
during its entire lifetime from 2017 to 2025.
Some of the properties of these seven pulsars are given in Table~\ref{tab:psr},
and their location within the $P$-$\Pdot$ diagram of pulsars is shown
in Figure~\ref{fig:ppdot}, where $P$ and $\Pdot$ are the pulsar spin period
and spin period time derivative, respectively, and
the relation of the former to pulsar spin frequency is $\nu=1/P$.

An outline of the paper is as follows.
A brief summary of each pulsar is given next in Section~\ref{sec:pulsarsum}.
Section~\ref{sec:data} describes the Chandra, NICER, and NuSTAR data
analysed in this work and their processing.
Section~\ref{sec:results} presents our results for each of the seven
pulsars, including their timing models.
Section~\ref{sec:discuss} summarizes the work presented here.

\subsection{Summary of pulsars} \label{sec:pulsarsum}

\psrzero\ is a relatively newly identified fast spinning young pulsar in
a supernova remnant and pulsar wind nebula in the Small Magellanic Cloud
\citep{owenetal11,maitraetal15}.
Its timing properties were first measured by \citet{maitraetal21},
and a rotation phase-connected timing model over an 8-month period
was presented in \citet{hoetal22}.
The pulsar has a narrow single peak pulse profile and high pulsed fraction
of $\approx 70$~percent in the 0.4--10~keV band.
Its rapid spin-down rate means \psrzero\ has the fourth highest measured
spin-down luminosity $\Edot$ among the  $\sim$3000 known pulsars.

\psrfive\ (also known as Big Glitcher) is the fastest spinning young pulsar,
is in the supernova remnant N157B in the Large Magellanic Cloud
\citep{wanggotthelf98,chenetal06}, and
has the the highest $\Edot$ among all known pulsars.
\psrfive\ has a narrow single peak pulse profile and X-ray pulsed fraction
of $\sim 20$~percent \citep{marshalletal98,kuiperhermsen15,hoetal20}.
While its spin frequency decreases
(by more than 0.16~Hz since 1999) over the more than 20 years
of combined observations (1999--2011 with RXTE and 2017--2025 with NICER),
a remarkable 68 glitches are measured, including 23 by NICER.
This yields the highest known average glitch rate of
$\sim 3.4\mbox{ yr$^{-1}$}$
\citep{marshalletal04,middleditchetal06,antonopoulouetal18,ferdmanetal18,hoetal20,hoetal22,abbottetal21a}.
Its glitches are unusual in their predictability, in particular
there is a correlation between the size of its glitches $\Delta\nu$
and time to its next glitch
\citep{middleditchetal06,antonopoulouetal18,ferdmanetal18,hoetal20,hoetal22}.
Its timing properties are also unusual, with a braking index
$n\equiv\nuddot\nu/\nudot^2=-1.25\pm0.01$ (1$\sigma$ error)
over the long-term (27-year duration) and
a value that approaches $\lesssim 7$ over the short-term
($\sim 100\mbox{ day}$) between glitches.

\psrzerofive\ is a young energetic pulsar (third highest $\Edot$)
in a pulsar wind nebula and supernova remnant in the Large Magellanic Cloud.
It is sometimes referred to as a twin of the Crab pulsar.
Its spin pulsations are detected in
radio \citep{manchesteretal93,johnstonromani03},
optical/UV \citep{middleditchpennypacker85,mignanietal19},
X-rays \citep{sewardetal84}, and gamma-rays \citep{fermi15}.
\psrzerofive\ undergoes small glitches and displays unusual
spin-down behavior, including
a large sudden change in $\nudot$ \citep{marshalletal15}
that led to an evolving braking index
(see, \citealt{espinozaetal24}, and references therein).

\psrfour\ (also known as Calvera) is a high Galactic latitude pulsar
\citep{zaneetal11,halpernetal13} residing within a diffuse radio ring that
seems to be a supernova remnant \citep{ariasetal22,rigosellietal24,grecoetal25}.
As a result, \psrfour\ appears to be a member of the class of neutron stars
known as central compact objects (see \citealt{deluca17}, for review).
If the distance of $\sim3\mbox{ kpc}$ determined from modeling the pulsar's
X-ray spectrum \citep{mereghettietal21} and the association between the
pulsar and supernova remnant candidate are correct,
then the age of the system is $\sim10-20\mbox{ kyr}$ \citep{grecoetal25}.
Monitoring observations of \psrfour\ during the first year of NICER in
2017--2018 yielded its phase-connected timing model \citep{bogdanovetal19},
and subsequent works gradually extended the timing model to over six years
using NICER data through 2023 November
\citep{mereghettietal21,hoetal22,rigosellietal24}.
Timing analysis by \citet{hoetal22} also yielded a pulsar position that
is inconsistent with the position and proper motion measurements made
by \citet{halperngotthelf15} using Chandra imaging data from 2007 and 2014.
Instead, the timing position and 2014 imaging position imply a proper motion
of $\mu_\alpha\cos\delta=+120\pm20\mbox{ mas yr$^{-1}$}$ and
$\mu_\delta=-3\pm20\mbox{ mas yr$^{-1}$}$.
A revised measurement of the position and proper motion
by \citet{rigosellietal24} using newer Chandra imaging data from 2024
found values that are much more consistent with the timing results of
\citet{hoetal22} (see also Section~\ref{sec:psrj1412}),
including a much more significant measurement of
$\mu_\alpha\cos\delta=+78.1\pm2.9\mbox{ mas yr$^{-1}$}$
and $\mu_\delta=+8.0\pm3.0\mbox{ mas yr$^{-1}$}$.

\psreightone\ is a young energetic pulsar in a pulsar wind nebula and
supernova remnant \citep{toriietal97,toriietal99,kaspietal01}.
It has spin pulsations that are only detected in X-ray up to 135~keV and
a high pulsed fraction of $>20$~percent
\citep{kuiperhermsen15,madsenetal20,zhengetal23,takataetal24}.

\psreightthree\ is a young highly energetic pulsar that produces a
pulsar wind nebula and is associated with the gamma-ray/TeV source
IGR~J18135$-$1751/HESS~J1813$-$178.
The pulsar's proper motion from the centre of the young supernova
remnant G12.82$-$0.02 implies an age of 1000--2200~yr \citep{dzibrodriguez21}.
\psreightthree\ has a broad single peak X-ray pulse profile
and high X-ray pulsed fraction
\citep{gotthelfhalpern09,halpernetal12,kuiperhermsen15,hoetal20a,takataetal24}.
Even though the pulsar has also been detected in radio,
its radio pulses suffer from extremely high scattering \citep{camiloetal21}.

\psreight\ is another young energetic pulsar that produces a
pulsar wind nebula and is associated with the gamma-ray/TeV source
IGR~J18490$-$0000/HESS~J1849$-$000.
The X-ray spectral properties of the pulsar and wind nebula have been
studied in, e.g., \citet{kuiperhermsen15,gagnonetal24,kimetal24,takataetal24}.
The pulsar has a broad single peak pulse profile and very high pulsed fraction 
\citep{kuiperhermsen15,bogdanovetal19,kimetal24,takataetal24}.
Phase-connected timing models have been derived using RXTE data
\citep{gotthelfetal11,kuiperhermsen15}
and Swift and NICER data \citep{bogdanovetal19,hoetal22,kimetal24}.

\section{Data analysis} \label{sec:data}

\begin{table}
\centering
\caption{Observation log}
\label{tab:data}
\begin{tabular}{llcc}
  \hline
Telescope & Pulsar & Observation date & Exposure \\
 & & & (ks) \\
  \hline
Chandra & \psrfive & 2002 Aug 23 & 28 \\
 & & 2002 Oct 29 & 20 \\
 & \psreightthree & 2012 Feb 12 & 20 \\
 & & 2021 Feb 10 & 20 \\
 & & 2021 Jun 23 & 20 \\
 & & 2024 Apr 3 & 21 \\
 & & 2024 Oct 7 & 20 \\
 & & 2025 Mar 23 & 20 \\
NICER & \psrzero & 2021 Jun 1--2024 Dec 1 & 609 \\
& \psrfive & 2017 Aug 17--2025 Jun 1 & 2171 \\
& \psrzerofive & 2017 Jul 25--2025 Jun 10 & 137 \\
& \psrfour & 2017 Sep 15--2025 Apr 23 & 1889 \\
& \psreightone & 2021 Sep 1--2025 May 23 & 317 \\
& \psreight & 2018 Feb 13--2025 Jun 8 & 474 \\
NuSTAR & \psrfive & 2016 Oct 17 & 105 \\
\hline
\end{tabular}
\end{table}

\subsection{Chandra data} \label{sec:chandra}

Chandra observed \psrfive\ twice in 2002 using the ACIS-S detector on
August 23 and October 29 (ObsID 2783) for a total of about 48~ks.
Chandra observed \psreightthree\ six times using the ACIS-S detector
in continuous clocking (CC) mode on 2012 February 12 (ObsID 12549),
2021 February 10 (ObsID 23545), 2021 June 23 (ObsID 23546),
2024 April 3 (ObsID 28355), 2024 October 7 (ObsID 28356), and
2025 March 23 (ObsID 28357)
for about 20~ks on each date (see Table~\ref{tab:data}).
We reprocess these data following the standard procedure using
\texttt{chandra\_repro} of the Chandra Interactive Analysis of
Observations (CIAO) package version 4.17 and Calibration Database
(CALDB) 4.12.0 \citep{fruscioneetal06}.
For data on \psrfive, we use \texttt{specextract} to extract source events
from a 1~arcsec radius circle centred on the pulsar position and with a
background region consisting of a 3~arcsec$\times$10~arcsec rectangular
box centred on the pulsar but excluding the source extraction circle
(see Figure~4 of \citealt{chenetal06}).
For data on \psreightthree, we extract source events from the
one-dimensional CC image along a 1.5~arcsec length
centred on the pulsar position and with a background region 11~arcsec
in length and about 10~arcsec from the source region;
we also consider a larger 3~arcsec extraction length to collect more
photons for spectral analysis (see Section~\ref{sec:psrj1813}),
although this gives somewhat weaker pulse detections.
Spectra are combined using \texttt{combine\_spectra} and binned using
\texttt{dmgroup} with a minimum of 25 counts per bin.

For timing analysis of \psreightthree, we select only events in the
2--8~keV energy range.
We transform the selected event time stamps from Terrestrial Time (TT)
to Barycentric Dynamical Time (TDB) using \texttt{axbary} and the pulsar
position.
Acceleration searches for the spin frequency are conducted using
PRESTO \citep{ransometal02} with a time bin of 2.85~ms.
Data are folded at the candidate pulse frequency using \texttt{prepfold},
and a refined frequency is determined.
Following the standard
procedure\footnote{\url{https://cxc.harvard.edu/ciao/threads/phase_bin}},
we compute rotation phases using \texttt{dmtcalc} and
create good time intervals (GTIs) using \texttt{dmgti}
for on-pulse emission,
i.e, photons with a rotation phase around the peak
of the pulse profile, and off-pulse emission,
which are photons with a phase outside the on-pulse phase range.
After correcting the exposure time, we extract phase resolved spectra using
\texttt{specextract}.
The off-pulse spectra can be used as the background for the on-pulse spectra,
such that the final pulsed spectra are the result of on minus off-pulse
emissions.

\subsection{NICER data} \label{sec:nicer}

For six of the seven pulsars considered in this work,
we use and report analyses of new NICER data,
which are summarized in Table~\ref{tab:data}.
For \psrzero, \psrzerofive, and \psreightone, searches for pulsations,
generation of pulse time-of-arrival (TOA), and determination of timing models
are performed following the procedures described in \citet{kuiperhermsen09}.
For \psrfive, \psrfour, and \psreight, their data are processed following
the same procedures described in \citet{hoetal22}, which we briefly
summarize here.  We refer the reader to \citet{hoetal22} for more details.
Note that NICER experienced a light leak\footnote{\url{https://heasarc.gsfc.nasa.gov/docs/nicer/analysis_threads/light-leak-overview}}
that affected data taken after 2023 May during orbit day.
As a result, the cadence of observations was reduced, and
fewer TOAs were obtained for pulsars such as \psrzero\ and \psrfour\ and
especially for \psrfive\ because of its high glitch activity. However, our
ability to obtain overall timing models was not severely impacted.

We process and filter NICER data using \texttt{nicerl2} of
HEASoft~6.22--6.35.2 \citep{heasarc14} and
NICERDAS~2018-03-01\_V003--2025-06-11\_V014.
While NICER is sensitive to 0.25--12~keV photons, we extract only events
within a particular energy range for each pulsar to optimize pulsation
searches
(1--7~keV for \psrfive, 0.37--1.97~keV for \psrfour, and
1.89--6~keV for \psreight).
We ignore time intervals of enhanced background affecting all detectors
by constructing a light curve binned at 16~s and removing intervals
strongly contaminated by background flaring when the count rate exceeds
a threshold value that is different for each pulsar
(10~c~s$^{-1}$ for \psrfive, 4.5~c~s$^{-1}$ for \psrfour, and
5~c~s$^{-1}$ for \psreight).
Using these filtering criteria, we obtain clean data for pulse timing analysis.
We use \texttt{barycorr} to transform between TT, used
for event time stamps, and TDB
and to account for effects of satellite motion with respect to the barycentre.
In all timing analyses performed here unless otherwise noted
(in particular, Sections~\ref{sec:psrj1412} and \ref{sec:psrj1849}),
source positions are held fixed at the values given in the corresponding
tables below, along with our adopted Solar system ephemeris.
As with analysis of Chandra data, we use PRESTO and \texttt{prepfold} to
perform a pulsation search and to determine the spin frequency of each pulsar.
A template pulse profile for each pulsar is produced by fitting a set
of NICER pulse profiles with a Gaussian shape; this template is then
used to determine the TOA of each observation following the
unbinned maximum likelihood technique described in \citet{rayetal11}.
Unless otherwise noted below, we use TEMPO2 \citep{hobbsetal06} to fit
TOAs with a timing model and to measure glitch parameters.
In particular, a pulsar's rotation phase $\phi$ is fit to a reference time $t_0$
based on a truncated Taylor series of the spin frequency and its time derivatives,
\begin{equation}
\phi(t) = \nu(t-t_0)+\frac{1}{2}\nudot(t-t_0)^2+\frac{1}{6}\nuddot(t-t_0)^3+\ldots.
\label{eq:phi}
\end{equation}
Glitches are fit to a model at the glitch epoch $\tg$ by
\begin{eqnarray}
\phi_{\rm g} &=& \Delta\phi+\Delta\nu(t-\tg)+\frac{1}{2}\Delta\nudot(t-\tg)^2
+\frac{1}{6}\Delta\nuddot(t-\tg)^3 \nonumber\\
&& +\Delta\nu_{\rm d}\tau_{\rm d}[1-e^{-(t-\tg)/\tau_{\rm d}}], \label{eq:phig}
\end{eqnarray}
where the last term accounts for a post-glitch recovery $\Delta\nu_{\rm d}$,
if present, that occurs on a timescale $\tau_{\rm d}$.

\subsection{NuSTAR data} \label{sec:nustar}

NuSTAR observed \psrfive\ on 2016 October 17 (ObsID 40201014002) for 105~ks.
We reprocess this data using \texttt{nupipeline} and \texttt{nuproducts}
of HEASoft~6.35.2 and NUSTARDAS~2025-03-11\_v2.1.5
and barycentre events with clock correction 20100101v211 and the
pulsar's position.
The source extraction region is a circle of 60~arcsec radius,
and the background region is a circle of
the same size and placed in a field with low counts.
For timing analysis, we use \texttt{xselect} to choose only events at
3--20~keV.
We use PRESTO and \texttt{prepfold} to perform pulsation searches on the
merged FPMA and FPMB event list
and to determine the spin frequency.
To extract pulsed spectra,
we first use \texttt{photonphase} in PINT \citep{luoetal21} and the spin
frequency to determine the rotation phase of each photon in the data.
We then use \texttt{maketime} to determine GTIs for on-pulse emission
and off-pulse emission.
Finally, we run \texttt{nuproducts} with these GTIs to produce on and
off-pulse spectra.

\section{Results} \label{sec:results}

\subsection{\psrzero} \label{sec:psrj0058}

\begin{figure}
\includegraphics[width=\columnwidth]{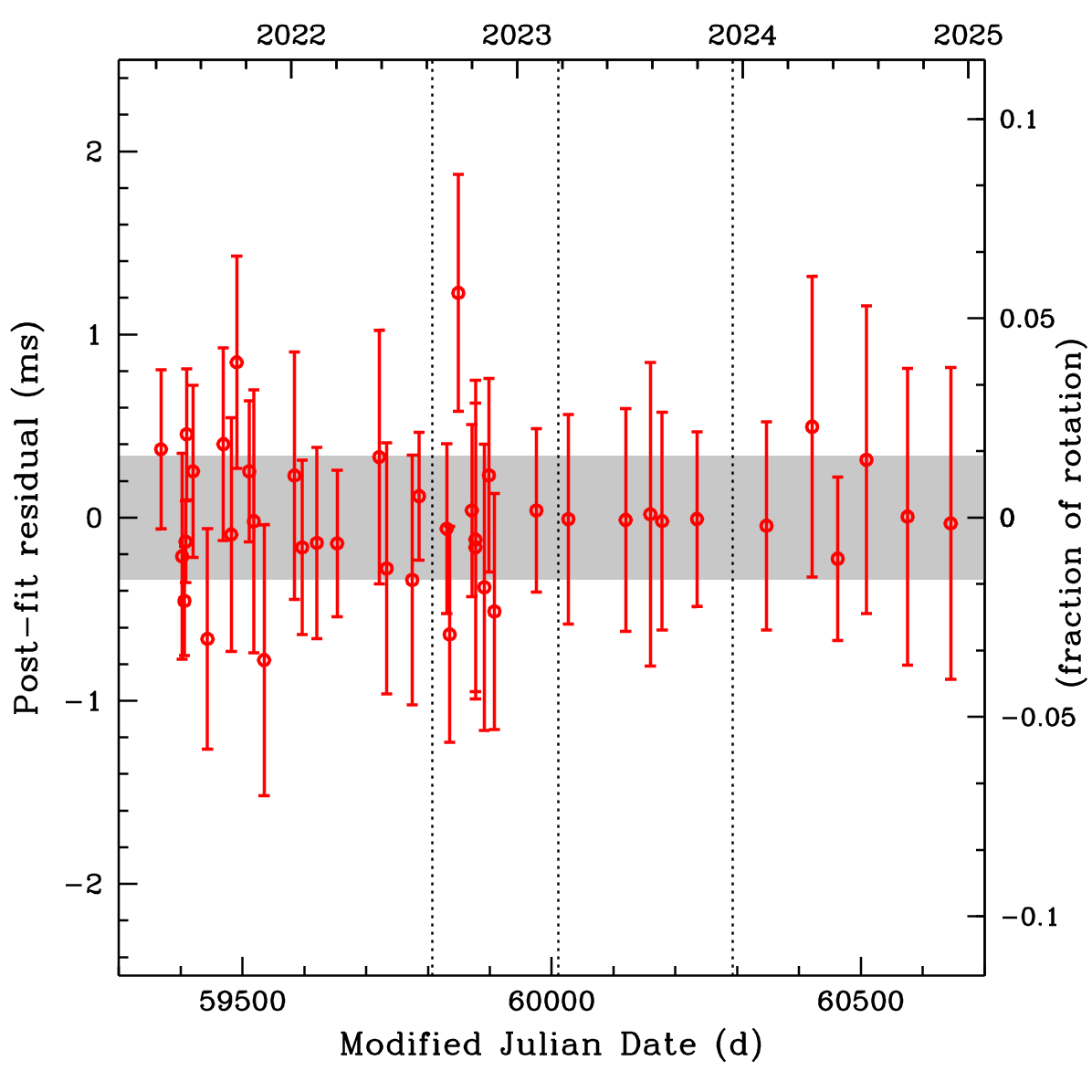}
\caption{Timing residuals of \psrzero\ from a best-fit of
NICER pulse times-of-arrival with the timing model given in
Table~\ref{tab:0058} and
the RMS residual illustrated by the shaded region between $\pm 0.339\mbox{ ms}$.
Errors are 1$\sigma$ uncertainty.
The vertical dotted lines indicate the approximate times of 3 spin-up glitches
(MJD 59807, 60011, 60293).}
\label{fig:0058}
\end{figure}

\begin{table}
\centering
\caption{Timing parameters of \psrzero.
Numbers in parentheses are 1$\sigma$ uncertainty in last digit.
The position is from a Chandra ACIS-S image (MJD 56332),
with a 90~percent confidence level uncertainty of 0.6~arcsec
\citep{maitraetal15}.}
\label{tab:0058}
\begin{tabular}{lc}
\hline
Parameter & Value \\
\hline
R.A. $\alpha$ (J2000) & $00^{\rm h}58^{\rm m}16\fs85$ \\
Decl. $\delta$ (J2000) & $-72\degr18\arcmin05\farcs60$ \\
Solar system ephemeris & DE405 \\
Range of dates (MJD) & 59368$-$60645 \\
Epoch (MJD TDB) & 59580 \\
$t_0$ (MJD) & 60026.9 \\
Frequency $\nu$ (Hz) & 45.93950855275(33) \\
Freq.\ 1st derivative $\nudot$ (Hz s$^{-1}$) & $-6.2263535(71)\times10^{-11}$ \\
Freq.\ 2nd derivative $\nuddot$ (Hz s$^{-2}$) & $3.962(29)\times10^{-21}$ \\
Glitch epoch 1 (MJD) & 59807$\pm$21 \\
$\Delta\phi_1$ & $0.125(28)$ \\
$\Delta\nu_1$ (Hz) & $7.35(72)\times10^{-8}$ \\
$\Delta\nudot_1$ (Hz s$^{-1}$) & $1.678(82)\times10^{-14}$ \\
Glitch epoch 2 (MJD) & 60011$\pm$11 \\
$\Delta\phi_2$ & $-0.029(53)$ \\
$\Delta\nu_2$ (Hz) & $3.6039(26)\times10^{-5}$ \\
$\Delta\nudot_2$ (Hz s$^{-1}$) & $-3.276(57)\times10^{-13}$ \\
$\Delta\nuddot_2$ (Hz s$^{-2}$) & $2.63(52)\times10^{-21}$ \\
Glitch epoch 3 (MJD) & 60293$\pm$48 \\
$\Delta\phi_3$ & $-0.02(16)$ \\
$\Delta\nu_3$ (Hz) & $3.1596(49)\times10^{-5}$ \\
$\Delta\nudot_3$ (Hz s$^{-1}$) & $-1.431(73)\times10^{-13}$ \\
$\Delta\nuddot_3$ (Hz s$^{-2}$) & $-1.70(57)\times10^{-21}$ \\
RMS residual ($\mu$s) & 338.9 \\
$\chi^2$/dof & 18.8/28 \\
Number of TOAs & 42 \\
\hline
\end{tabular}
\end{table}

NICER observations of \psrzero\ began on 2021 June 1.
\citet{hoetal22} reported a phase-connected timing model through
2022 January 25.
Here we extend the timing model by about 3 years with NICER data through
2024 December 1, for a total timespan of 3.5 years.
Note that additional NICER data through 2025 June 1 are not of sufficient
quality to enable detection of the spin pulsations of \psrzero.
Figure~\ref{fig:0058} shows residuals of our best-fit timing model,
which is given in Table~\ref{tab:0058}.

Our regular cadence of observations of \psrzero\ reveal that this young
pulsar underwent three glitches during 3.5~years of monitoring.
Not only does this suggest a high glitch rate for \psrzero,
but two of the three glitches are large,
i.e., $\Delta\nu\gtrsim10\mbox{ $\mu$Hz}$ (see, e.g., \citealt{fuentesetal17}).
Both these properties are typical for young pulsars, including \psrfive\
and the Vela pulsar \citep{antonopoulouetal22}.
Meanwhile the small glitch $\Delta\nu_1$ is accompanied by a positive change
in spin-down rate, i.e., $\Delta\nudot_1>0$, which is contrary to typical
behavior of glitches and could indicate timing noise similar to that seen in
\psrzerofive\ (see \citealt{espinozaetal24} and Section~\ref{sec:psrb0540}).

\subsection{\psrfive} \label{sec:psrj0537}

\begin{table*}
\centering
\caption{Timing parameters of \psrfive.
Columns are interglitch segment number, segment start and end dates,
timing model epoch, spin frequency and its first two time derivatives,
interglitch braking index $\nig (=\nuddot\nu/\nudot^2)$,
timing model residual, goodness-of-fit measure,
and number of times of arrival.
Numbers in parentheses are 1$\sigma$ uncertainty in last digit.
Values for segments 1--7, 8--10, and 11--14 are similar to those given in
\citet{hoetal20}, \citet{abbottetal21a}, and \citet{hoetal22}, respectively.
Position of R.A.~$=05^{\rm h}37^{\rm m}47^{\rm s}\!\!.416$,
decl.~$=-69^\circ10\arcmin19\farcs88$ (J2000) is
from a Chandra ACIS-I image (MJD 51442), with 1$\sigma$ uncertainty
of $\sim0.6$~arcsec \citep{townsleyetal06}.
Solar system ephemeris used is DE421.}
\label{tab:0537}
\begin{tabular}{cccccccccccc}
\hline
Segment & Start & End & Epoch & $t_0$ & $\nu$ & $\nudot$ & $\nuddot$ & $\nig$ & RMS & $\chi^2/\mbox{dof}$ & TOAs \\
& (MJD) & (MJD) & (MJD) & (MJD) & (Hz) & ($10^{-10}\mbox{ Hz s$^{-1}$}$) & ($10^{-20}\mbox{ Hz s$^{-2}$}$) & & ($\mu$s) & & \\
\hline
 0 & 57984 & 58058 & 58020 & 58049.1 & 61.9243742602(16) & $-$1.995468(50)  & [1]$^{a}$ &       --- & 66.1 & 3.14/1 & 4 \\
 1 & 58108 & 58142 & 58124 & 58138.0 & 61.9225972030(30) & $-$1.99611(30)   & [1]$^{a}$ &       --- & 1.29 & ---$^b$ & 3 \\
 2 & 58162 & 58349 & 58255 & 58268.4 & 61.9203729962(9)  & $-$1.9969767(20) & 0.564(12) &  8.75(18) & 124  & 74.5/17 & 21 \\
 3 & 58377 & 58423 & 58399 & 58402.8 & 61.9178962511(30) & $-$1.996890(24)  & 5.95(64)  & 92.3(99)  & 56.4 & 8.43/7 & 11 \\
 4 & 58426 & 58561 & 58493 & 58482.1 & 61.9162995588(37) & $-$1.9972611(85) & 0.813(82) & 12.6(13)  & 251  & 183/9 & 13 \\
 5 & 58571 & 58630 & 58600 & 58601.6 & 61.9144624211(18) & $-$1.997402(26)  & [1]$^{a}$ &      ---  & 98.5 & 13.1/4 & 7 \\
 6 & 58645 & 58805 & 58723 & 58720.0 & 61.9123666199(15) & $-$1.9972887(36) & 0.880(29) & 13.66(44) & 143  & 70.8/12 & 16 \\
 7 & 58810 & 58863 & 58836 & 58835.0 & 61.9104242102(61) & $-$1.997420(39)  & 5.76(89)  & 89(13)    & 81.9 & 9.34/2 & 6 \\
 8 & 58872 & 58991 & 58931 & 58931.0 & 61.9088087390(25) & $-$1.9975354(70) & 1.058(81) & 16.4(13)  & 174  & 129/13 & 17 \\
 9 & 58995 & 59047 & 59020 & 59018.6 & 61.9072733762(21) & $-$1.996986(40)  & [1]$^{a}$ &       --- & 148  & 53.7/8 & 11 \\
10 & 59050 & 59099 & 59074 & 59071.3 & 61.9063499483(47) & $-$1.997617(20)  & 3.64(81)  & 56(13)    & 60.9 & 8.95/6 & 10 \\
11 & 59108 & 59283 & 59195 & 59195.2 & 61.9042953587(11) & $-$1.9975882(25) & 0.595(17) & 9.24(26)  & 180  & 229/27 & 31 \\
12 & 59286 & 59350 & 59318 & 59318.2 & 61.9021804218(40) & $-$1.997892(16)  & 0.71(38)  & 11.1(60)  & 82.4 & 15.4/6 & 10 \\
13 & 59352 & 59447 & 59399 & 59401.0 & 61.9007944153(27) & $-$1.9978015(93) & 1.63(13)  & 25.3(21)  & 132  & 60/13 & 17 \\
14 & 59461 & 59514 & 59487 & 59485.8 & 61.8992920385(27) & $-$1.998284(32)  & [1]$^{a}$ &       --- & 72.7 & 9.06/3 & 6 \\
15 & 59529 & 59687 & 59607 & 59612.6 & 61.8972423813(11) & $-$1.9979145(33) & 0.798(23) & 12.38(36) & 152  & 124/23 & 27 \\
16 & 59696 & 59736 & 59716 & 59717.0 & 61.8953668356(27) & $-$1.998020(45)  & [1]$^{a}$ &       --- & 98.6 & 21.6/9 & 12 \\
17 & 59745 & 59882 & 59813 & 59805.7 & 61.8937075333(16) & $-$1.9978974(51) & 1.566(42) & 24.13(65) & 162  & 109/18 & 22 \\
18 & 59889 & 60027 & 59958 & 59957.1 & 61.8912289268(14) & $-$1.9981564(29) & 0.846(31) & 13.11(48) & 140  & 132/22 & 26 \\
19 & 60070 & 60221 & 60145 & 60135.6 & 61.8880344209(9)  & $-$1.9982364(18) & 0.582(16) &  9.03(25) & 73.8 & 16.9/13 & 17 \\
20 & 60224 & 60358 & 60291 & 60294.2 & 61.8855349303(12) & $-$1.9983872(23) & 0.864(30) & 13.39(46) & 69.2 & 15.1/11 & 15 \\
21 & 60401 & 60585 & 60492 & 60502.7 & 61.8820960295(51) & $-$1.9981842(48) & 0.781(51) & 12.10(79) & 126  & 16.6/2 & 6 \\
22 & 60600 & 60754 & 60686 & 60666.4 & 61.8787773846(19) & $-$1.9987440(57) & 0.749(38) & 11.59(58) & 135 & 38.5/6 & 10 \\
23 & 60771 & 60827 & 60799 & 60803.1 & 61.8768369527(25) & $-$1.998925(40)  & [1]$^{a}$ &       --- & 129 & 22.8/5 & 8 \\
\hline
\multicolumn{11}{l}{$^a$\,$\nuddot$ is fixed at $10^{-20}\mbox{ Hz s$^{-2}$}$
due to low number of TOAs and/or timespan too short to measure $\nuddot$.
$^b$No fit performed.}
\end{tabular}
\end{table*}

\begin{table*}
\centering
\caption{Glitch parameters of \psrfive.  Columns are glitch number and epoch
and change in rotation phase, changes in spin frequency and its
first two time derivatives, and post-glitch recovery $\Delta\nu_{\rm d}$
and $\tau_{\rm d}$ at each glitch.
Numbers in parentheses are 1$\sigma$ uncertainty in last digit.
Values for glitches 1--7, 8--10, and 11--15 are consistent with those given in
\citet{hoetal20}, \citet{abbottetal21a}, and \citet{hoetal22}, respectively.}
\label{tab:0537glitch}
\begin{tabular}{cccccccc}
\hline
Glitch & Glitch epoch & $\Delta\phi$ & $\Delta\nu$ & $\Delta\nudot$ & $\Delta\nuddot$ & $\Delta\nu_{\rm d}$ & $\tau_{\rm d}$ \\
& (MJD) & (cycle) & ($\mu$Hz) & ($10^{-13}\mbox{ Hz s$^{-1}$}$) & ($10^{-20}\mbox{ Hz s$^{-2}$}$) & ($\mu$Hz) & (d) \\
\hline
 1 & 58083 $\pm$ 25 & $-$0.016(8)  & 16.132(2)   & $-$1.54(5)   & --- & --- & ---\\
 2 & 58152 $\pm$ 11 &    0.23(16)  & 36.026(44)  & $-$1.48(20)  & --- & 0.532(86) & 6.3(25) \\
 3 & 58363 $\pm$ 14 &    0.167(50) &  7.829(55)  & $-$2.29(36)  &    5.4(11) & --- & ---\\
 4 & 58424 $\pm$ 2  & $-$0.35(23)  & 25.33(28)   & $-$2.1(12)   & $-$5.1(20) & --- & ---\\
 5 & 58566 $\pm$ 5  & $-$0.324(21) &  9.205(16)  & $-$0.890(51) & --- & --- & ---\\
 6 & 58637 $\pm$ 8  &    0.029(19) & 26.986(13)  & $-$0.861(39) & $-$0.120(27)$^a$ & --- & ---\\
 7 & 58807 $\pm$ 3  &    0.310(22) &  7.565(30)  & $-$2.21(26)  &    4.88(95)  & --- & ---\\
 8 & 58868 $\pm$ 5  & $-$0.05(17)  & 23.95(13)   & $-$2.20(48)  & $-$5.23(77) & 0.48(83) & 5.1(28) \\
 9 & 58993 $\pm$ 3  &    0.06(12)  &  0.41(10)   & $-$0.26(84)  & --- & --- & ---\\
10 & 59049 $\pm$ 3  & $-$0.209(11) &  8.424(12)  & $-$1.074(50) & --- & --- & ---\\
11 & 59103 $\pm$ 5  &    0.39(37)  & 33.81(32)   & $-$1.26(83)  & $-$3.15(94) & 0.2(13) & 9.2(29) \\
12 & 59285 $\pm$ 2  & $-$0.256(11) & 7.8722(81)  & $-$0.938(26) & --- & --- & ---\\
13 & 59351 $\pm$ 2  &    0.513(20) & 12.272(34)  & $-$0.79(22)  &    0.92(49) & --- & ---\\
14 & 59454 $\pm$ 8  &    0.309(14) & 16.604(10)  & $-$1.712(38) & --- & --- & ---\\
15 & 59522 $\pm$ 8  & $-$0.294(28) & 22.062(16)  & $-$0.519(46) & $-$0.202(23) & --- & ---\\
16 & 59692 $\pm$ 5  & $-$0.175(16) &  5.835(14)  & $-$0.857(61) & --- & --- & ---\\
17 & 59741 $\pm$ 5  & $-$0.420(19) & 15.400(16)  & $-$1.062(69) &    0.556(37) & --- & ---\\
18 & 59886 $\pm$ 4  &    0.103(7)  & 24.3746(49) & $-$1.767(19) & $-$0.710(30) & --- & ---\\
19 & 60028 $\pm$ 1  &    1.97(23)  & 34.00(10)   & $-$1.02(16)  & $-$0.39(12) & 0.25(25) & 29.7(46) \\
20 & 60223 $\pm$ 1  &    0.224(7)  & 21.0960(21) & $-$0.706(23) & --- & 0.282(30) & 33.1(52) \\
21 & 60379 $\pm$ 22 &    0.386(60) & 31.494(29)  & $-$1.223(85) & $-$0.130(78) & --- & ---\\
22 & 60592 $\pm$ 8  & $-$0.244(32) & 31.042(15)  & $-$1.842(67) & $-$0.033(61) & --- & ---\\
23 & 60762 $\pm$ 9  & $-$0.352(18) & 10.918(13)  & $-$0.909(43) & --- & --- & ---\\
\hline
\multicolumn{8}{l}{$^a$In the preceding segment, $\nuddot$ is fixed at
$10^{-20}\mbox{ Hz s$^{-2}$}$ due to a low number of TOAs
(see Table~\ref{tab:data}).  $\Delta\nuddot$ is the} \\
\multicolumn{8}{l}{difference between this fixed value and $\nuddot$
in the next segment.}
\end{tabular}
\end{table*}

\begin{figure*}
\includegraphics[width=1.8\columnwidth]{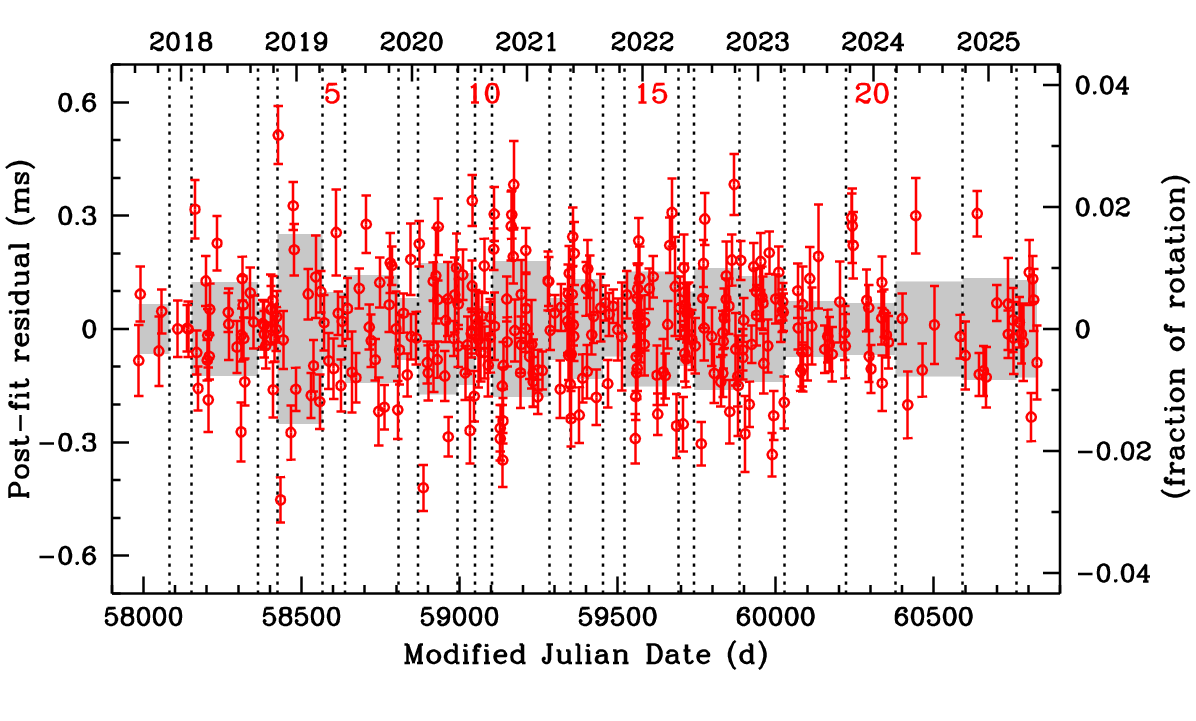}
\caption{Timing residuals of \psrfive\ from a best-fit of
NICER pulse times-of-arrival with the timing model given in
Table~\ref{tab:0537} and
RMS residuals illustrated by the shaded regions.
Errors are 1$\sigma$ uncertainty.
Segments are labelled by numbers and separated by the occurrence
of a glitch, each of which is denoted by a vertical dotted line.}
\label{fig:0537}
\end{figure*}

\begin{figure}
\includegraphics[width=\columnwidth]{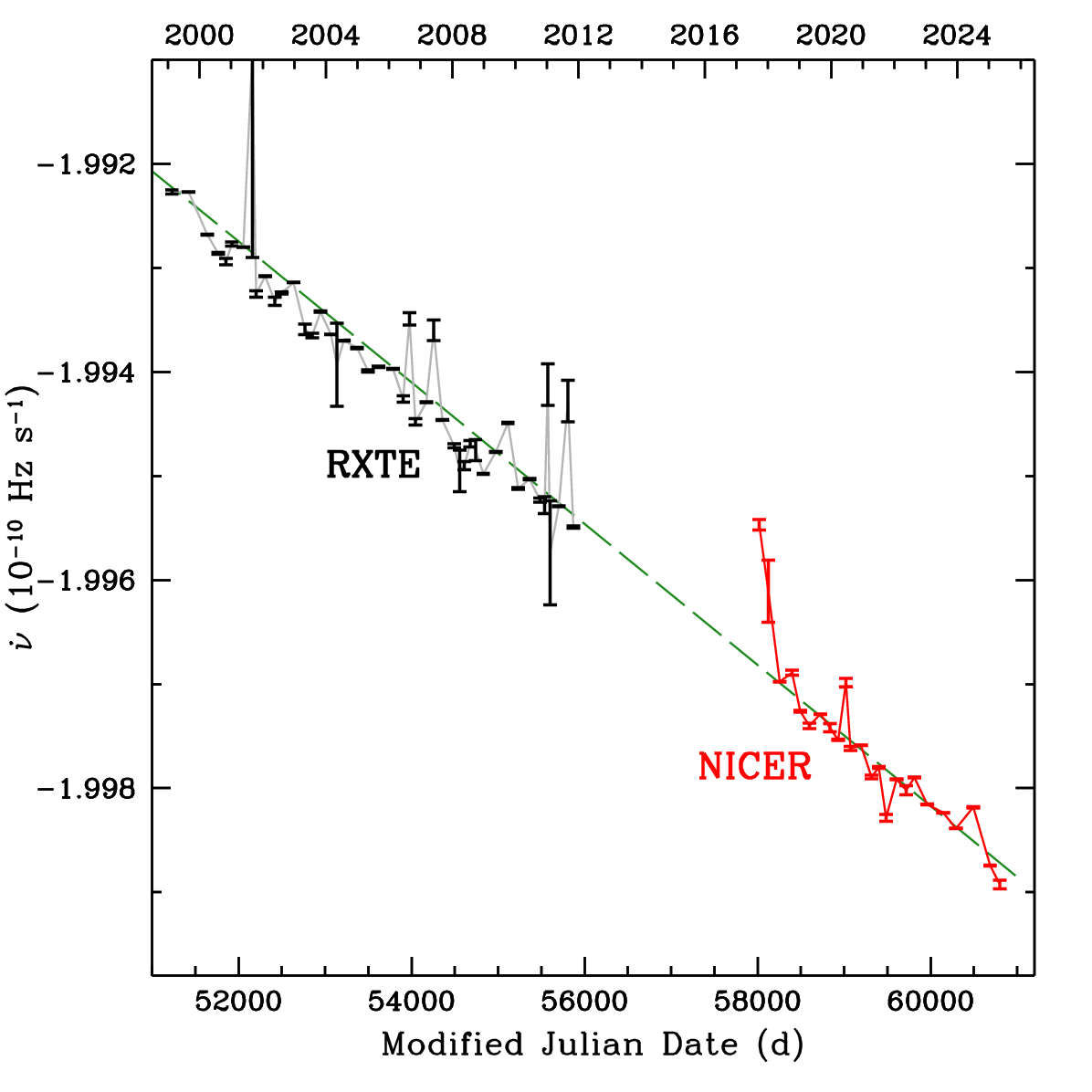}
\caption{Evolution of spin frequency time derivative $\nudot$ of \psrfive,
where $\nudot$ is from the timing model of each segment
(see Table~\ref{tab:0537} for NICER and
Table~1 of \citealt{antonopoulouetal18} for RXTE).
Errors are 1$\sigma$ uncertainty.
Dashed line shows a linear fit of NICER and RXTE data with
best-fit $\nuddot=-7.85\times 10^{-22}\mbox{ Hz s$^{-2}$}$.}
\label{fig:0537f1}
\end{figure}

\begin{figure}
\includegraphics[width=\columnwidth]{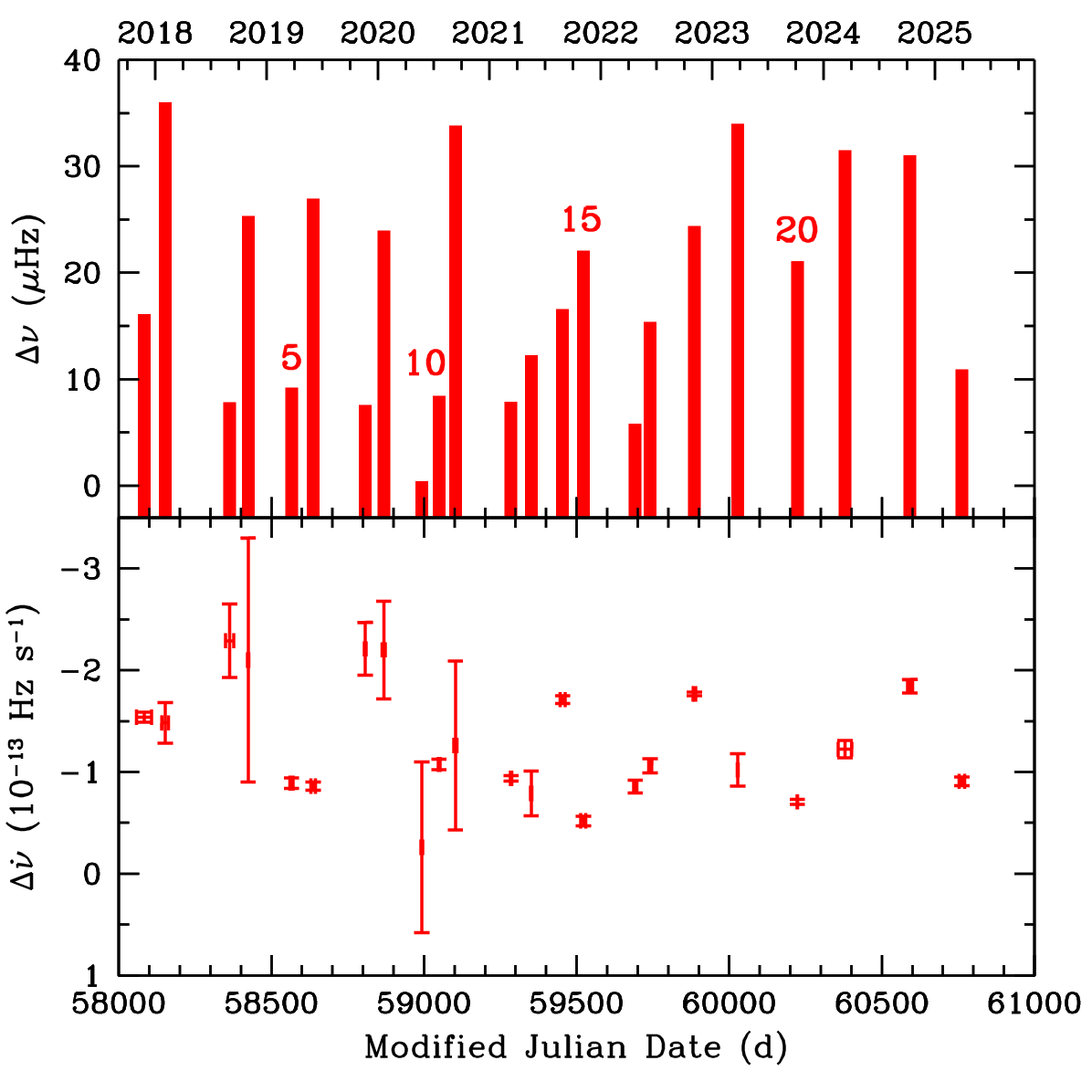}
\caption{
Glitch $\Delta\nu$ (top) and $\Delta\nudot$ (bottom) as functions of time.
Errors in $\Delta\nudot$ are 1$\sigma$ uncertainty.}
\label{fig:0537glitch}
\end{figure}

Because of the fast spin of \psrfive\ and that its spin pulsations are
only detectable in X-ray, few instruments have been able to track its
timing properties over long periods of time.
RXTE performed high-cadence observations of \psrfive\ from 1999 to 2011,
and these data yielded phase-connected timing models during this 13-year
period \citep{antonopoulouetal18,ferdmanetal18}.
NICER started regular monitoring observations of \psrfive\ soon after
its launch in mid-2017.
Phase-connected timing models and 8 glitches measured with NICER data
through 2020 April were presented in \citet{hoetal20}.
Subsequent timing models through 2020 October, including 3 further glitches,
and then through 2022 February, including 4 more glitches, were reported
in \citet{abbottetal21a} and \citet{hoetal22}, respectively.
Here we report timing models using the remaining NICER data from 2022 March
through 2025 June, with 8 new glitches detected during this period
(first 6 of which are also reported in \citealt{zubietaetal26}).
Tables~\ref{tab:0537} and \ref{tab:0537glitch} show timing models and
glitch parameters derived from the complete NICER dataset from 2017 to 2025,
including those given in our previous works above.
We use our previous naming convention, where each segment is separated
by a glitch and is labelled by glitch number, with segment 1 occurring
after glitch 1, which is the first NICER-detected glitch.
Figure~\ref{fig:0537} shows residuals of our best-fit timing models to
all the NICER data,
Figure~\ref{fig:0537f1} plots the spin-down rate $\nudot$ measured by
RXTE and NICER since 1999, and
Figure~\ref{fig:0537glitch} displays glitch parameters $\Delta\nu$ and
$\Delta\nudot$ for the 23 glitches measured using NICER.
A best-fit linear decline for the long-term spin-down rate yields
$\nuddot=(-7.85\pm0.04)\times 10^{-22}\mbox{ Hz s$^{-2}$}$ and a braking index
$n=-1.222\pm0.007$
calculated using $\nu$ and $\nudot$ from around the middle
of the 25~years of RXTE and NICER observations.
For a few glitches, an exponential recovery term is included in the glitch
model [see equation~(\ref{eq:phig})],
but the presence of the recovery is not considered when characterizing the
intervals between glitches.
A detailed study of exponential recoveries of glitches of \psrfive\ is
conducted in \citet{zubietaetal26}.

\begin{figure}
\includegraphics[width=\columnwidth]{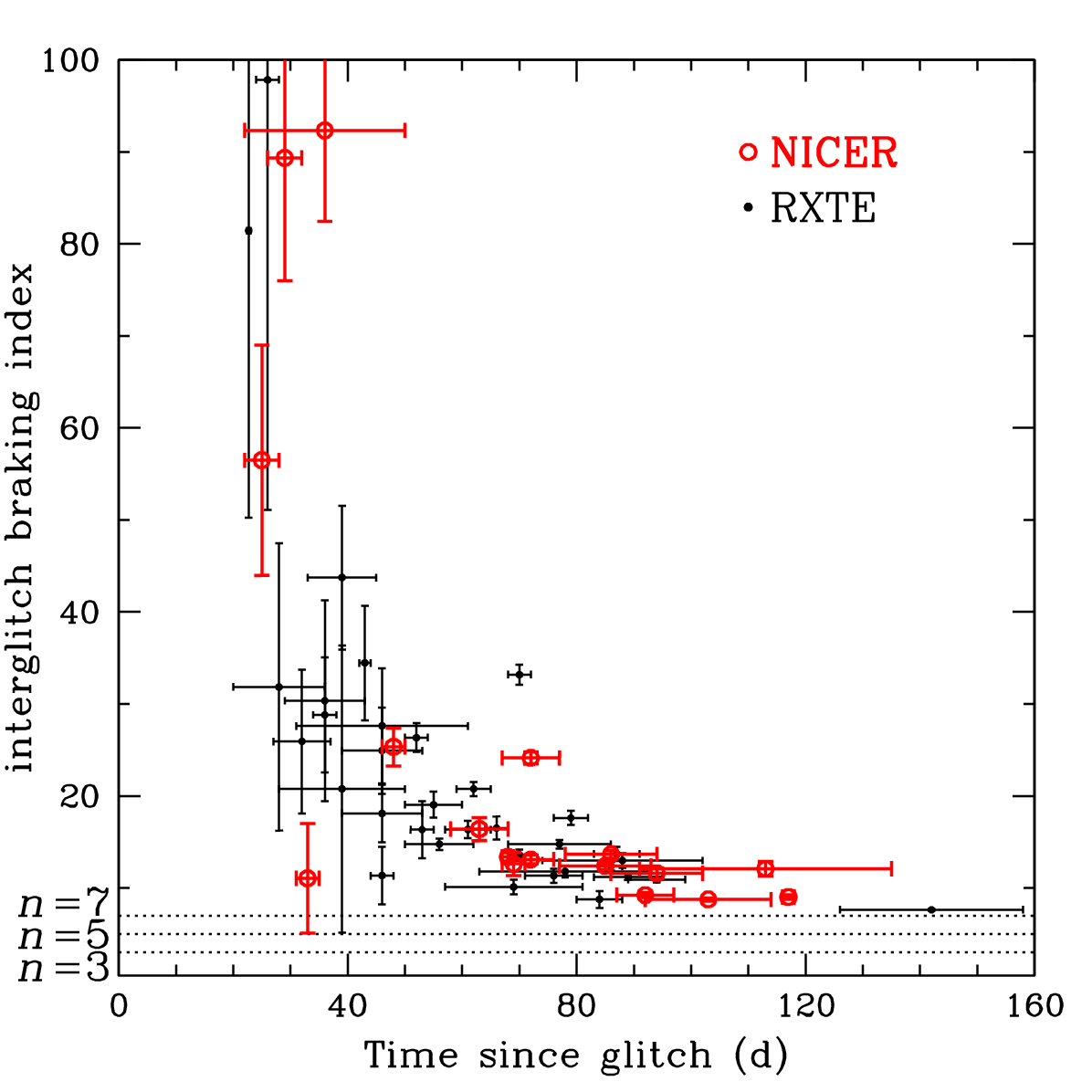}
\caption{
Interglitch braking index $n_{\rm ig}$ of \psrfive\ calculated
from spin parameters of each segment between glitches as a
function of time since the last glitch.
Large and small circles denote NICER and RXTE values, respectively
(from Tables~\ref{tab:0537} and \ref{tab:0537glitch} and
\citealt{antonopoulouetal18}).
Errors in $n_{\rm ig}$ are 1$\sigma$.
Horizontal dotted lines indicate braking index $n=3$, 5, and 7,
which are expected for pulsar spin-down by electromagnetic dipole
radiation, gravitational wave-emitting mountain, and
gravitational wave-emitting r-mode oscillation, respectively.}
\label{fig:0537nig}
\end{figure}

As discussed in \citet[][and references therein]{hoetal22},
the short-term spin-down behavior of \psrfive\ (i.e., behavior between glitches)
is very different from the long-term behavior seen in Figure~\ref{fig:0537f1}.
In particular, the interglitch braking index $\nig$,
calculated using the spin parameters of each segment between
glitches (see Table~\ref{tab:0537}), is non-negative
and much greater than the canonical value of 3 that is predicted for
conventional spin-down by electromagnetic dipole radiation at constant
magnetic field and moment of inertia.
This is illustrated in Figure~\ref{fig:0537nig}, which shows
$\nig$ measured using NICER and RXTE, with the latter values taken
from \citet{antonopoulouetal18},
and time since last glitch is the epoch of the segment minus the epoch
of the corresponding glitch
(e.g., time since glitch 23 $=60794-60762=32\mbox{ d}$).
It is clear that large values of $\nig$ are measured for short
times after a glitch
and that small $\nig$ are measured after long post-glitch times.
In other words, it appears there is recovery back to a rotational
behavior that is characterized by a braking index $\lesssim7$
after disruption by a glitch.
Fits to an exponential decay yield decay timescales of 19--44~d,
with a longer timescale for a lower asymptotic braking index \citep{hoetal20}.
Braking indices of 5 and 7 are expected for spin-down by gravitational
wave quadrupole and r-mode emission, respectively.

\begin{figure}
\includegraphics[width=\columnwidth]{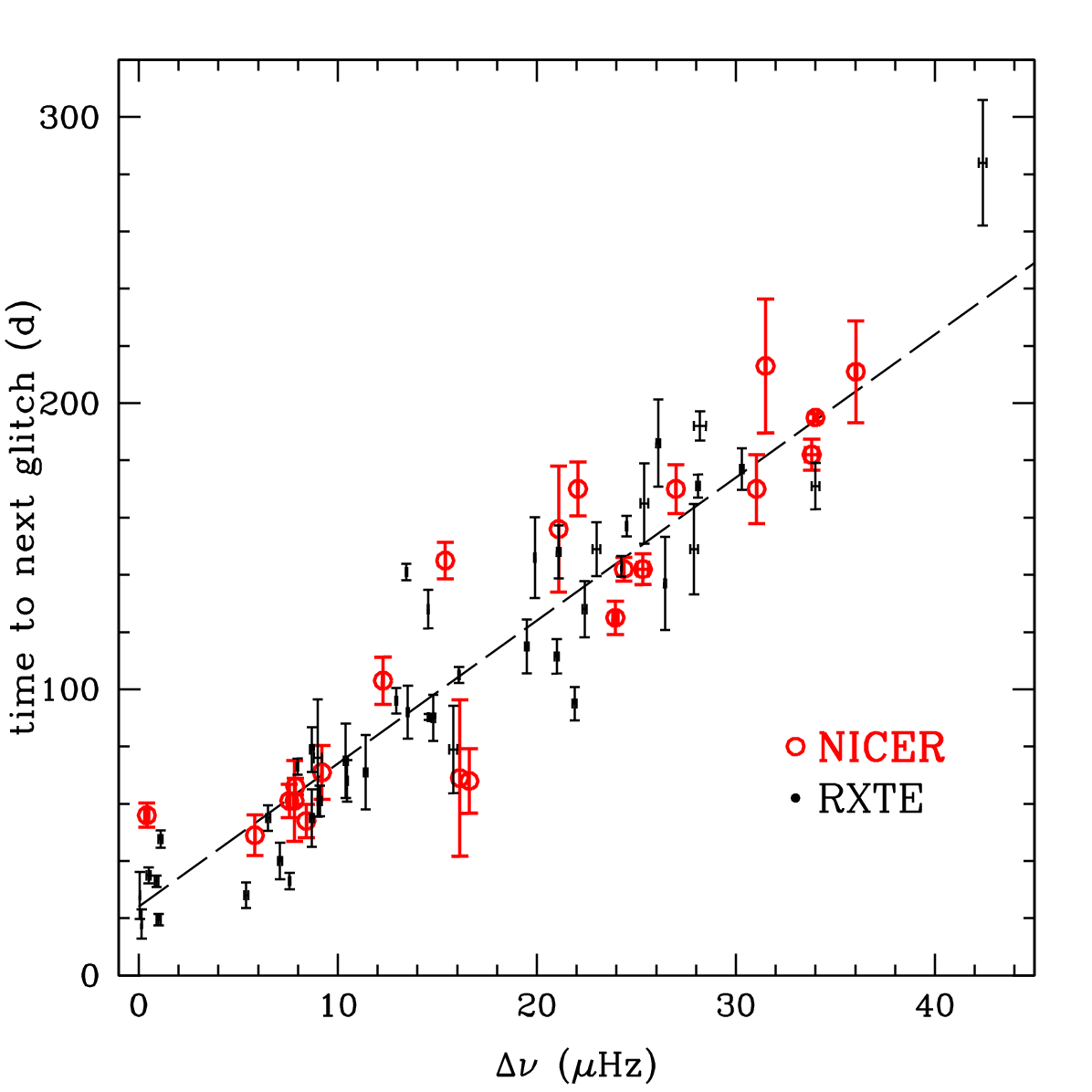}
\caption{
Correlation between time interval to
the next glitch $\Delta T$ and size of glitch $\Delta\nu$ of \psrfive.
Large and small circles denote NICER and RXTE values, respectively
(from Table~\ref{tab:0537glitch} and \citealt{antonopoulouetal18}).
Errors in $\Delta\nu$ are 1$\sigma$.
Dashed line shows linear fit result
$\Delta T=50\mbox{ d }(\Delta\nu/\mbox{10 $\mu$Hz})+24\mbox{ d}$.}
\label{fig:0537tg}
\end{figure}

Glitches of \psrfive\ are unique in that the time to next glitch is
correlated with the size of the preceding glitch
(see also references in Section~\ref{sec:pulsarsum}).
This is illustrated in Figure~\ref{fig:0537tg}.
The correlation can be fit by time to
\mbox{next glitch $=50\mbox{ d }(\Delta\nu/\mbox{10 $\mu$Hz})+24\mbox{ d}$},
in agreement with that found in \citet{hoetal20,hoetal22}.
This correlation enables prediction of when glitches will occur in \psrfive.

Because of the non-imaging wide field-of-view of NICER and bright X-ray
emission from the surrounding region around \psrfive, we do not perform
spectral analyses of the NICER data here and leave this for future work.
The X-ray spectrum of the pulsar has been studied previously.
\citet[][see also \citealt{townsleyetal06}]{chenetal06} fit the total
(pulsed plus unpulsed or on plus off-pulse) Chandra spectrum of \psrfive\
with an absorbed power law model and found a hydrogen absorption column
density $\NH=5.6^{+0.5}_{-0.3}\times10^{21}\mbox{ cm$^{-2}$}$,
power law index $\Gamma=1.73^{+0.11}_{-0.06}$, and unabsorbed 0.5--10~keV
flux of $1.9\times10^{-12}\mbox{ erg s$^{-1}$ cm$^{-2}$}$,
where uncertainties are at 90~percent confidence.
Meanwhile, \citet{kuiperhermsen15} derived the 0.7--250~keV pulsed flux
spectrum of \psrfive\ using RXTE and XMM-Newton data and by assuming an absorbed
curved power law spectral model and modeling the pulse profiles as a
function of energy.

\begin{figure}
\includegraphics[width=\columnwidth]{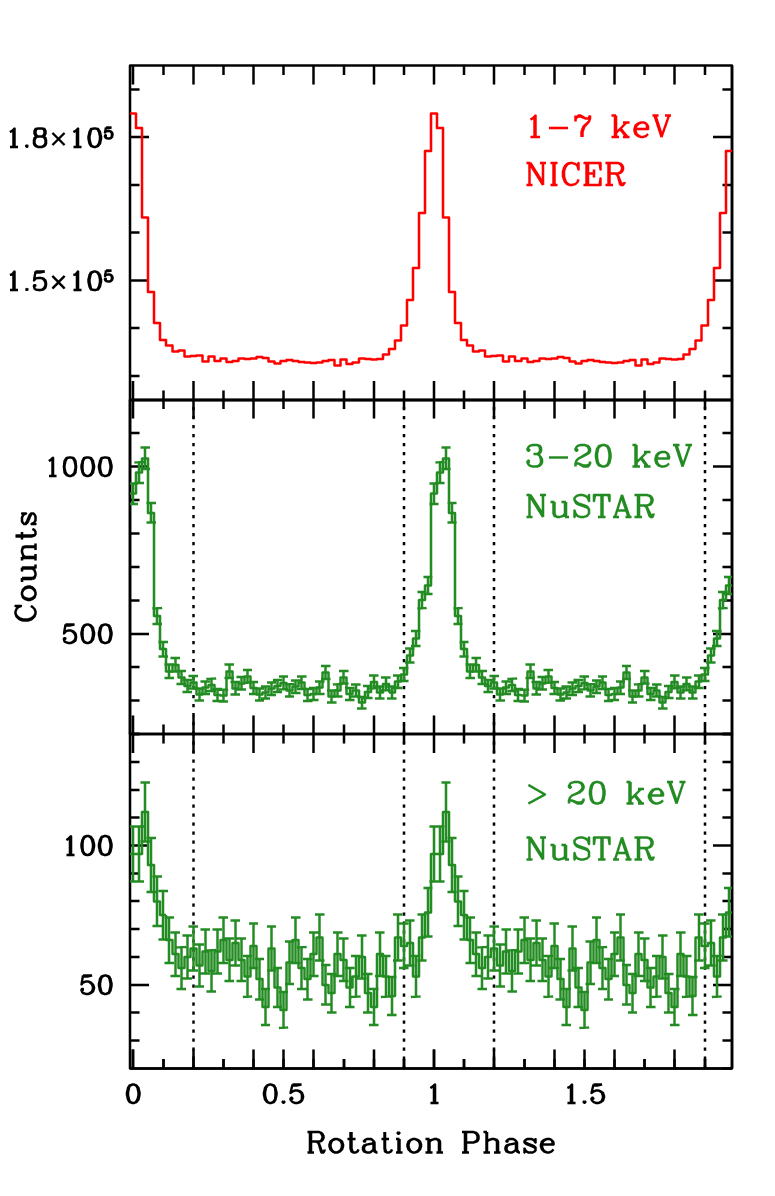}
\caption{Pulse profile of \psrfive\ from NICER data at 1--7~keV (top)
and NuSTAR data at 3--20~keV (middle) and $>20$~keV (bottom).
Errors are 1$\sigma$ uncertainty.
Two rotation cycles are shown for clarity.
Phase alignment between NICER and NuSTAR is arbitrary since datasets
do not overlap.
Dotted lines indicate phase ranges chosen for on-pulse (phase~$<0.2$ or $>0.9$)
and off-pulse ($0.2<\mbox{phase}<0.9$) NuSTAR spectra.
\label{fig:0537_pp}}
\end{figure}

Here we provide a more straightforward measurement of the pulsed spectrum
at X-ray energies 3--79~keV using a 105~ks NuSTAR observation
from 2016 (see Table~\ref{tab:data}).
Note that \citet{bambaetal22} analysed this same NuSTAR data, alongside
Suzaku data, to determine the total (pulsar+pulsar wind nebula) spectrum 
at 1.5--79~keV but did not perform timing analysis to extract the pulsed
spectrum of \psrfive.
First, we follow the procedure described in Section~\ref{sec:nustar} and
measure a spin frequency $\nu=61.93022931(7)\mbox{ Hz}$ at MJD~57678.64,
where the $1\sigma$ uncertainty in the last digit is given in parentheses.
Figure~\ref{fig:0537_pp} shows the resulting 3--20 and $>$20~keV
pulse profiles when the data are folded at this frequency, alongside
the 1--7~keV pulse profile measured using NICER data.
This spin frequency is $\sim300\mbox{ $\mu$Hz}$ higher than the frequency
predicted assuming the approximate pulsar spin-down rate during the time
since the last RXTE data five years prior to the NuSTAR observation
(see Figure~\ref{fig:0537f1}).
Such a deviation can be expected, given that \psrfive\ likely underwent
$>15$ spin-up glitches during this time.
Analogously, the NuSTAR frequency is $\sim30\mbox{ $\mu$Hz}$ lower than the
frequency extrapolated from the first NICER data.  This is consistent with
a few spin-up glitches during the intervening year between NuSTAR and NICER
observations.

To extract pulsed spectra, we determine peak on-pulse emission as rotation
phases below 0.2 or above 0.9 and off-pulse emission as phases outside
the on-pulse ranges (see Figure~\ref{fig:0537_pp}).
The off-pulse spectra are used as the background for the on-pulse spectra,
such that the final pulsed spectra are the result of on minus off.
We model the final pulsed spectra in Xspec \citep{arnaud96} using an absorbed
power law, in particular \texttt{tbabs} with abundances from \citet{wilmsetal00}
and cross-sections from \citet{verneretal96} and \texttt{powerlaw}.
We fix the absorption $\NH$ to the value $7\times10^{21}\mbox{ cm$^{-2}$}$,
based on our analysis below of the same Chandra data in \citet{chenetal06},
where they determined the slightly lower value noted above.
We find a best-fit $\Gamma=1.62\pm0.05$ with $\chi^2/\mbox{dof}=164/165$
and unabsorbed
2--10~keV flux of $(2.3\pm0.1)\times10^{-12}\mbox{ erg s$^{-1}$ cm$^{-2}$}$ and
3--79~keV flux of $(7.8\pm0.4)\times10^{-12}\mbox{ erg s$^{-1}$ cm$^{-2}$}$.
Accounting for the on-pulse phase width of 0.3, the phase-averaged
unabsorbed pulsed fluxes are
$(6.9\pm0.3)\times10^{-13}\mbox{ erg s$^{-1}$ cm$^{-2}$}$ for 2--10~keV and
$(2.3\pm0.1)\times10^{-12}\mbox{ erg s$^{-1}$ cm$^{-2}$}$ for 3--79~keV.

\begin{figure}
\includegraphics[width=\columnwidth]{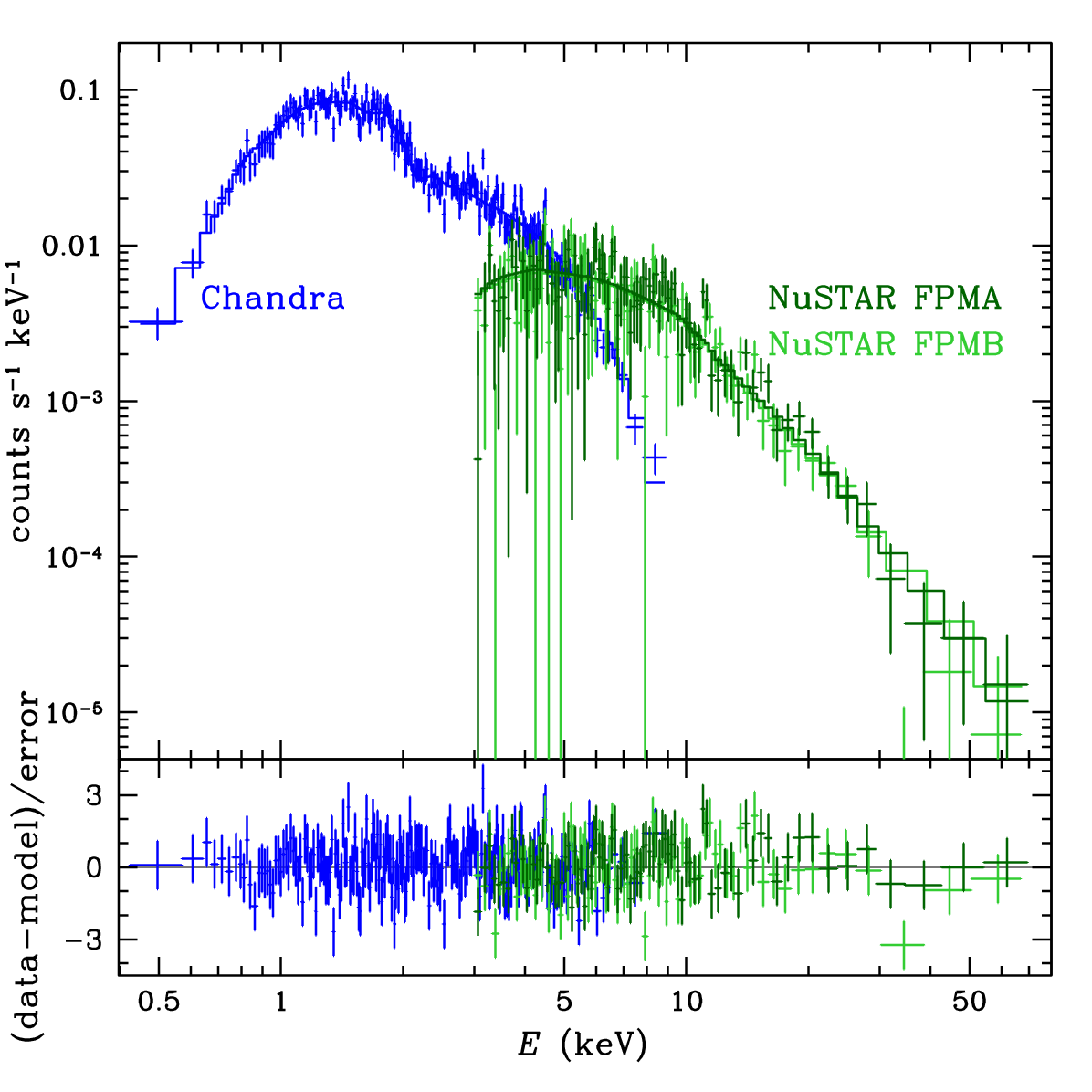}
\caption{Pulsed spectra of \psrfive\ from Chandra and NuSTAR FPMA and FPMB data.
Top panel shows data with 1$\sigma$ errors (crosses) and power law
spectral model (solid lines).
Bottom panel shows $\chi^2=\mbox{(data-model)/error}$.
\label{fig:0537_spectra2}}
\end{figure}

Since the best-fit power law index $\Gamma\approx1.6$ for the pulsed
component at 3--79~keV is similar to the $\Gamma\approx1.7$ obtained
by \citet{chenetal06} for the pulsed$+$unpulsed component at 0.5-10~keV,
we test whether a single power law could be used to describe the
broadband ($\sim$~0.3--79~keV) pulsed spectrum of \psrfive.
We use the same Chandra observation (see Table~\ref{tab:data}) and
source and background regions to extract the pulsed$+$unpulsed spectrum
as that used in \citet{chenetal06}.
For a joint spectral modeling of both the Chandra and NuSTAR data, we
include an extra model fit parameter, \texttt{constant}, in addition to
\texttt{tbabs} and \texttt{powerlaw}.
This normalization parameter is allowed to vary for the Chandra spectrum
but is fixed to 1 for the NuSTAR spectra.
The variation of this normalization from unity for the Chandra spectrum can
be interpreted as a combination of
the contribution of the unpulsed component to the Chandra spectrum,
assuming this contribution is energy-independent, and
detector calibration uncertainties between Chandra and NuSTAR.
We find a best-fit with $\chi^2/\mbox{dof}=344/356$ that yields
a Chandra normalization of $0.73\pm0.04$,
$\NH=(7.1\pm0.3)\times10^{21}\mbox{ cm$^{-2}$}$, and $\Gamma=1.67\pm0.03$.
The phase-averaged unabsorbed 2--10~keV pulse flux is
$(7.1\pm0.2)\times10^{-13}\mbox{ erg s$^{-1}$ cm$^{-2}$}$,
which is consistent with the fluxes derived from ASCA and RXTE data by
\citet{marshalletal98} and RXTE and XMM-Newton data by \citet{kuiperhermsen15}.
Figure~\ref{fig:0537_spectra2} shows the NuSTAR pulsed spectra and the
``derived'' Chandra pulsed spectra, as well as our best-fit model.

\subsection{\psrzerofive} \label{sec:psrb0540}

\begin{figure*}
\includegraphics[width=1.8\columnwidth]{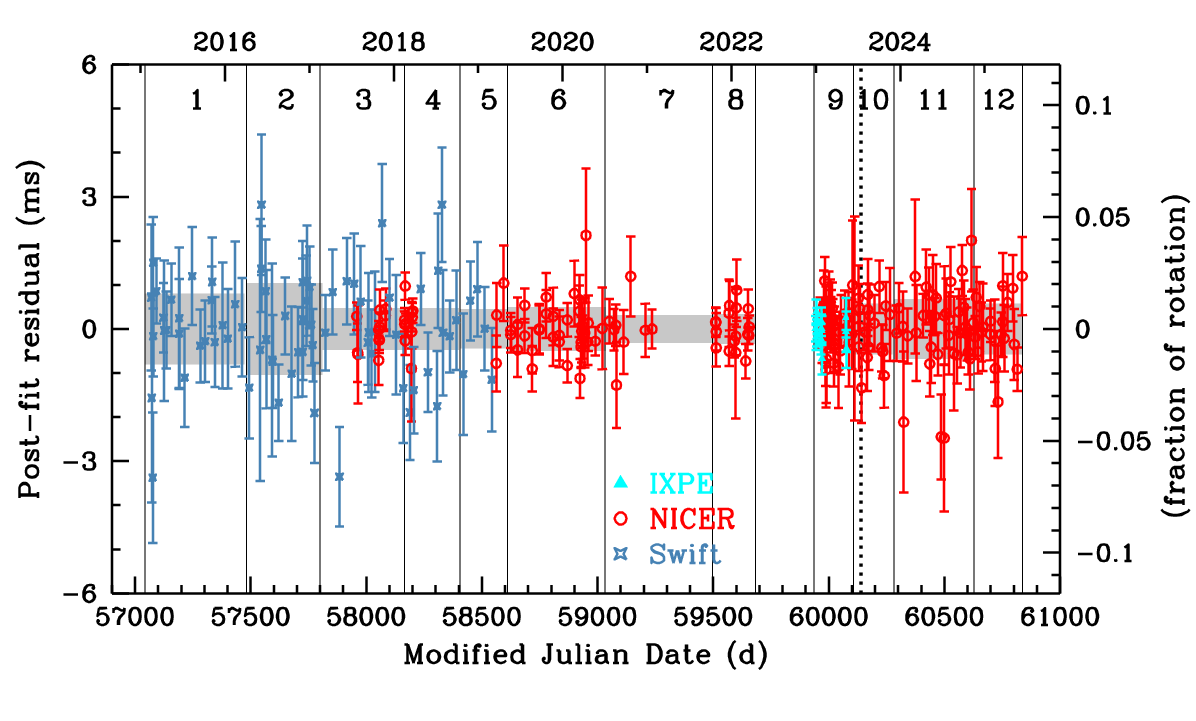}
\caption{Timing residuals of \psrzerofive\ from a best-fit of
IXPE (triangles), NICER (circles), and Swift (crosses) pulse
times-of-arrival with the timing model given in
Table~\ref{tab:0540} and RMS residuals illustrated by the shaded regions.
Errors are 1$\sigma$ uncertainty.
Segments are denoted by vertical solid lines and labelled by numbers;
overlaps of some segments are not indicated.
Vertical dotted line at MJD~60139 denotes the occurrence of a timing event
(see text for details).}
\label{fig:0540}
\end{figure*}

\begin{table*}
\centering
\caption{Timing parameters of \psrzerofive.
Columns are similar to those in Table~\ref{tab:0537}.
Numbers in parentheses are 1$\sigma$ uncertainty in last digit.
Position of R.A.~$=05^{\rm h}40^{\rm m}11\fs202$,
decl.~$=-69\degr19\arcmin54\farcs17$ is from Hubble WFPC2 images (MJD 54272),
with a 1$\sigma$ uncertainty of 0.07~arcsec \citep{mignanietal10}.
Solar system ephemeris used is DE405.}
\label{tab:0540}
\begin{tabular}{cccccccccccc}
\hline
Segment & Start & End & Epoch & $t_0$ & $\nu$ & $\nudot$ & $\nuddot$ & $n$ & RMS & $\chi^2/\mbox{dof}$ & TOAs \\
& (MJD) & (MJD) & (MJD) & (MJD) & (Hz) & ($10^{-10}\mbox{ Hz s$^{-1}$}$) & ($10^{-21}\mbox{ Hz s$^{-2}$}$) & & ($\mu$s) & & \\
\hline
 1 & 57044 & 57473 & 57430 & 57247.5 & 19.6930733748(23) & $-$2.5286496(38) & 0.090(26) & 0.0277(80) & 811 & 14.0/19 & 23 \\
 2 & 57482 & 57810 & 57620 & 57649.1 & 19.6889224954(17) & $-$2.5284950(28) & 0.805(95) & 0.248(29) & 1043 & 18.9/17 & 21 \\
 3 & 57800 & 58163 & 58000 & 57976.1 & 19.68062121257(74) & $-$2.5282227(14) & 2.041(41) & 0.628(13) & 475 & 25.3/21 & 25 \\
 4 & 58165 & 58400 & 58300 & 58266.7 & 19.6740689746(14) & $-$2.5274940(69) & 2.58(15) & 0.795(47) & 481 & 27.5/16 & 20 \\
 5 & 58406 & 58881 & 58650 & 58654.0 & 19.66642738628(57) & $-$2.52646402(47) & 3.485(13) & 1.0736(40) & 448 & 9.75/13 & 17 \\
 6 & 58611 & 59036 & 58830 & 58834.9 & 19.66249864699(53) & $-$2.52594021(57) & 3.255(16) & 1.0031(49) & 495 & 21.5/20 & 24 \\
 7 & 59033 & 59550 & 59280 & 59235.8 & 19.6526803925(19) & $-$2.52455830(52) & 3.798(30) & 1.1710(93) & 313 & 3.93/6 & 10 \\
 8 & 59497 & 59683 & 59610 & 59591.3 & 19.6454838675(17) & $-$2.5234484(99) & 3.52(43) & 1.09(13) & 342 & 12.9/11 & 15 \\
 9 & 59935 & 60107 & 60020 & 60021.6 & 19.63654744724(73) & $-$2.5219243(20) & 4.58(15) & 1.414(46) & 345 & 33.6/44 & 48 \\
10 & 60107 & 60276 & 60200 & 60197.5 & 19.6326257165(19) & $-$2.5213478(85) & 5.99(54) & 1.85(17) & 577 & 15.1/14 & 18 \\
11 & 60282 & 60630 & 60470 & 60458.3 & 19.62674530613(83) & $-$2.52021134(79) & 4.559(34) & 1.409(10) & 683 & 33.9/31 & 35 \\
12 & 60628 & 60836 & 60730 & 60731.9 & 19.6210851128(15) & $-$2.5190981(23) & 5.25(16) & 1.623(50) & 577 & 15.9/16 & 20 \\
\hline
\end{tabular}
\end{table*}

NICER observations of \psrzerofive\ began on 2017 July 25.
Previous works reported timing models using NICER data from a relatively
short time range of 2023 January 19 to May 3, supplemented by IXPE data
\citep{xieetal24}, and
from 2022 November 18 to 2023 December 25 \citep{tuoetal24}.
In \citet{espinozaetal24}, we presented results from our timing models
using NICER data from the start of its observations on 2017 July 25
through 2024 July 29, supplemented by IXPE and Swift data.
Here we provide the timing models and update these models to include
all the effective NICER data through 2025 June 10.
Figure~\ref{fig:0540} shows timing residuals of the TOAs used to obtain
our best-fit timing models, which are given in Table~\ref{tab:0540}.
Here the timing model segments for \psrzerofive\ are not separated
by glitches, in contrast to the glitch-separated segments in the timing
model for, e.g., \psrfive\ given in Section~\ref{sec:psrj0537}.
The segments are determined by periods over which a model that includes
$\nuddot$ provides a resonable fit to the data.

\begin{figure}
\includegraphics[width=\columnwidth]{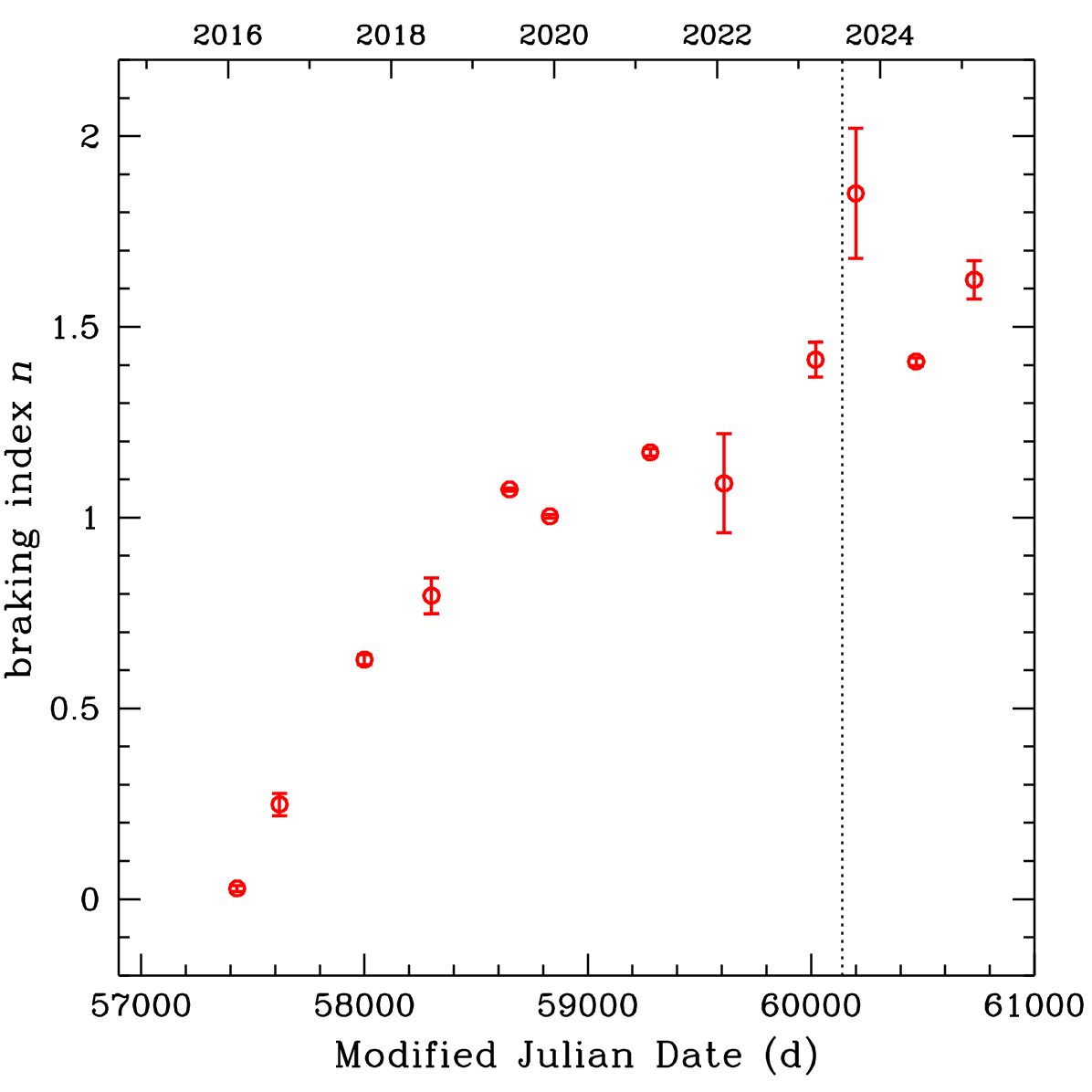}
\caption{Evolution of braking index $n=\nuddot\nu/\nudot^2$ of \psrzerofive\ as
measured during each segment of the timing model given in Table~\ref{tab:0540}.
Errors are 1$\sigma$ uncertainty.
Vertical dotted line at MJD~60139 denotes the occurrence of a timing event
(see text for details).}
\label{fig:0540n}
\end{figure}

The latest data and longer term timing analysis continues to indicate
the timing event that occurred around MJD~60139 was not a spin-down glitch
as reported by \citet{tuoetal24} but a more subtle change in spin-down
behavior (see \citealt{espinozaetal24}, for more details).
Figure~\ref{fig:0540n} shows the braking index $n$
($=\nuddot\nu/\nudot^2$), calculated for each segment
of the timing model given in Table~\ref{tab:0540}, as a function of time.
The slowly growing braking index measured by
\citet{espinozaetal24} through early 2024 continues through 2025.
We also see the effect of the timing event on the braking index measurement
near that time.

\subsection{\psrfour} \label{sec:psrj1412}

\begin{figure}
\includegraphics[width=\columnwidth]{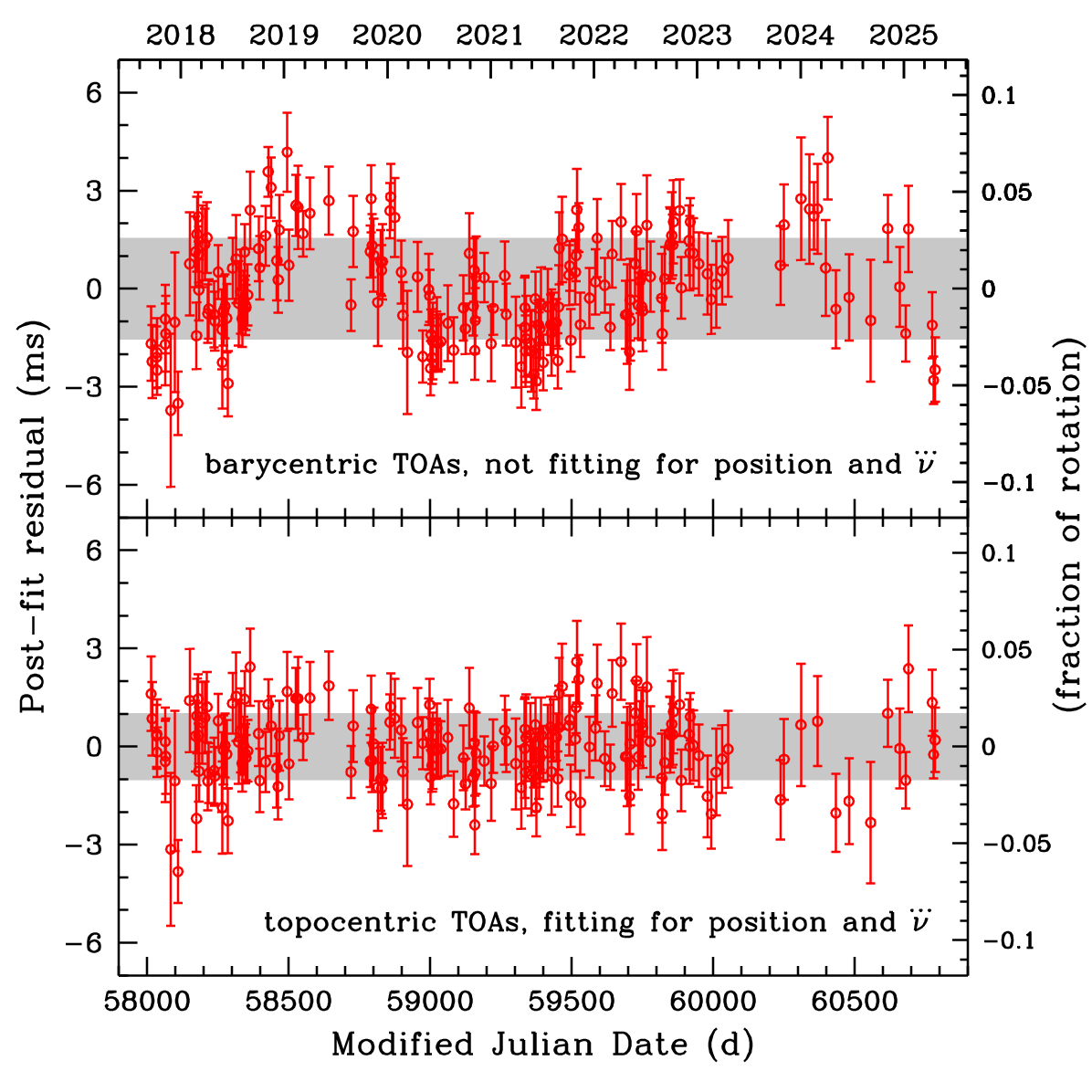}
\caption{Top panel shows timing residuals of \psrfour\ from a best-fit
of NICER barycentric pulse times-of-arrivals with a timing model that
includes $\nuddot$ but not $\nudddot$ nor position.
Bottom panel shows timing residuals of \psrfour\ from a best-fit
of NICER topocentric TOAs with the timing model given in
Table~\ref{tab:1412}.
Errors are 1$\sigma$ uncertainty.
The RMS residuals are illustrated by the shaded regions between
$\pm1.55\mbox{ ms}$ (top) and between $\pm1.03\mbox{ ms}$ (bottom).}
\label{fig:1412}
\end{figure}

\begin{table}
\centering
\caption{Timing parameters of \psrfour.
Numbers in parentheses are 1$\sigma$ uncertainty in last digit.
Position epoch is MJD 58750.
Proper motion is not taken into account in the fit.}
\label{tab:1412}
\begin{tabular}{lc}
\hline
Parameter & Value \\
\hline
R.A. $\alpha$ (J2000) & $14^{\rm h}12^{\rm m}56\fs035(19)$ \\
Decl. $\delta$ (J2000) & $+79\degr22\arcmin03\farcs518(43)$ \\
Solar system ephemeris & DE405 \\
Range of dates (MJD) & 58014$-$60784 \\
Epoch (MJD TDB) & 59400 \\
$t_0$ (MJD) & 58815.3 \\
Frequency $\nu$ (Hz) & 16.89205540623(5) \\
Freq.\ 1st derivative $\nudot$ (Hz s$^{-1}$) & $-9.420861(24)\times10^{-13}$ \\
Freq.\ 2nd derivative $\nuddot$ (Hz s$^{-2}$) & $-2.6597(33)\times10^{-23}$ \\
Freq.\ 3rd derivative $\nudddot$ (Hz s$^{-3}$) & $2.52(22)\times10^{-32}$ \\
RMS residual (ms) & 1.030 \\
$\chi^2$/dof & 201.8/182 \\
Number of TOAs & 189 \\
\hline
\end{tabular}
\end{table}

NICER observations of \psrfour\ began on 2017 September 15, and
\citet{bogdanovetal19} reported a phase-connected timing model
using the first year of data through 2018 October.
\citet{mereghettietal21} extended the timing model to 3.4~yr
using NICER data through 2021 February, and
\citet{hoetal22} further extended the model another year using data
through 2022 February.
More recently, \citet{rigosellietal24} reported the 6.1~yr timing model
using data through 2023 November.
Here we complete the timing model using the remaining effective NICER
data through 2025 April 23, covering a period of over 7.6~yr.

To construct a timing model that includes the pulsar's position as a
fit parameter, as done in \citet{hoetal22}, we calculate 189 spacecraft
topocentric, not barycentric, TOAs from the dataset, and we fit these
TOAs using PINT.
In our timing analysis, we do not include the pulsar's proper motion,
as determined by \citet{rigosellietal24} using Chandra data,
so that we can obtain a position independent of their imaging analysis.
The bottom panel of Figure~\ref{fig:1412} shows timing residuals of the
topocentric TOAs used to obtain
our best-fit timing model, which is given in Table~\ref{tab:1412}.
For comparison, we show in the top panel timing residuals from a
TEMPO2 best-fit of the corresponding barycentric TOAs to a timing
model that includes $\nu$, $\nudot$, and $\nuddot$ but not $\nudddot$
and position;
the worse fit and long timescale variations of the residuals are evident.
We find that our position from timing is consistent with the Chandra
astrometric imaging position from \citet{rigosellietal24} of
R.A.~$=14^{\rm h}12^{\rm m}56\fs126(8)$ and
decl.~$=+79\degr22\arcmin03\farcs54(2)$ at MJD 60374,
as well as the pulsar's proper motion (see Section~\ref{sec:pulsarsum}),
and our uncertainty is about twice that of Chandra's.

\subsection{\psreightone} \label{sec:psrj1811}

\begin{figure}
\includegraphics[width=\columnwidth]{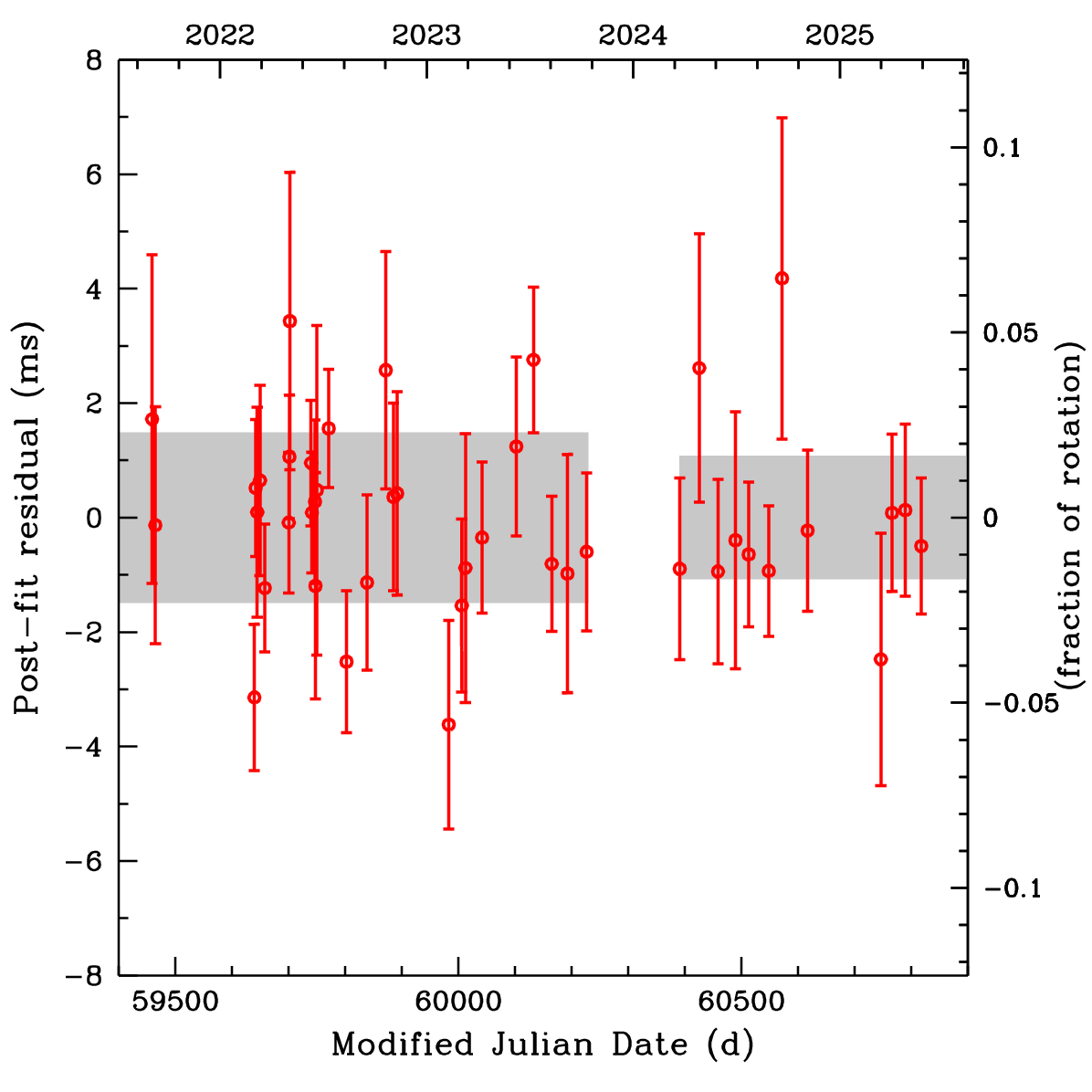}
\caption{Timing residuals of \psreightone\ from two separate best-fits of
NICER pulse times-of-arrival with the timing models given in
Table~\ref{tab:1811}.
Errors are 1$\sigma$ uncertainty.
The RMS residual of each best-fit is illustrated by the shaded regions
between $\pm 1.49\mbox{ ms}$ before MJD~60230 and
between $\pm 1.08\mbox{ ms}$ after MJD~60390.}
\label{fig:1811}
\end{figure}

\begin{table}
\centering
\caption{Timing parameters of \psreightone.
Numbers in parentheses are 1$\sigma$ uncertainty in last digit.
The position is from a Chandra ACIS-S image (MJD 51762),
with a 1$\sigma$ uncertainty of 0.6~arcsec \citep{kaspietal01}.}
\label{tab:1811}
\begin{tabular}{lc}
\hline
Parameter & Value \\
\hline
R.A. $\alpha$ (J2000) & $18^{\rm h}11^{\rm m}29\fs22$ \\
Decl. $\delta$ (J2000) & $-19\degr25\arcmin27\farcs6$ \\
Solar system ephemeris & DE405 \\
Range of dates (MJD) & 59459$-$60230 \\
Epoch (MJD TDB) & 59701 \\
$t_0$ (MJD) & 59871.5 \\
Frequency $\nu$ (Hz) & 15.4544586828(11) \\
Freq.\ 1st derivative $\nudot$ (Hz s$^{-1}$) & $-1.0669215(75)\times10^{-11}$ \\
Freq.\ 2nd derivative $\nuddot$ (Hz s$^{-2}$) & $-2.64(16)\times10^{-22}$ \\
Freq.\ 3rd derivative $\nudddot$ (Hz s$^{-3}$) & $-5.6(12)\times10^{-30}$ \\
RMS residual (ms) & 1.49 \\
$\chi^2$/dof & 31.8/25 \\
Number of TOAs & 30 \\
\hline
Range of dates (MJD) & 60390$-$60818 \\
Epoch (MJD TDB) & 60458 \\
$t_0$ (MJD) & 60512.8 \\
Frequency $\nu$ (Hz) & 15.4537601837(19) \\
Freq.\ 1st derivative $\nudot$ (Hz s$^{-1}$) & $-1.068594(86)\times10^{-11}$ \\
Freq.\ 2nd derivative $\nuddot$ (Hz s$^{-2}$) & $-8.7(18)\times10^{-22}$ \\
Freq.\ 3rd derivative $\nudddot$ (Hz s$^{-3}$) & $5.2(13)\times10^{-29}$ \\
RMS residual (ms) & 1.08 \\
$\chi^2$/dof & 5.92/7 \\
Number of TOAs & 12 \\
\hline
\end{tabular}
\end{table}

\begin{figure}
\includegraphics[width=\columnwidth]{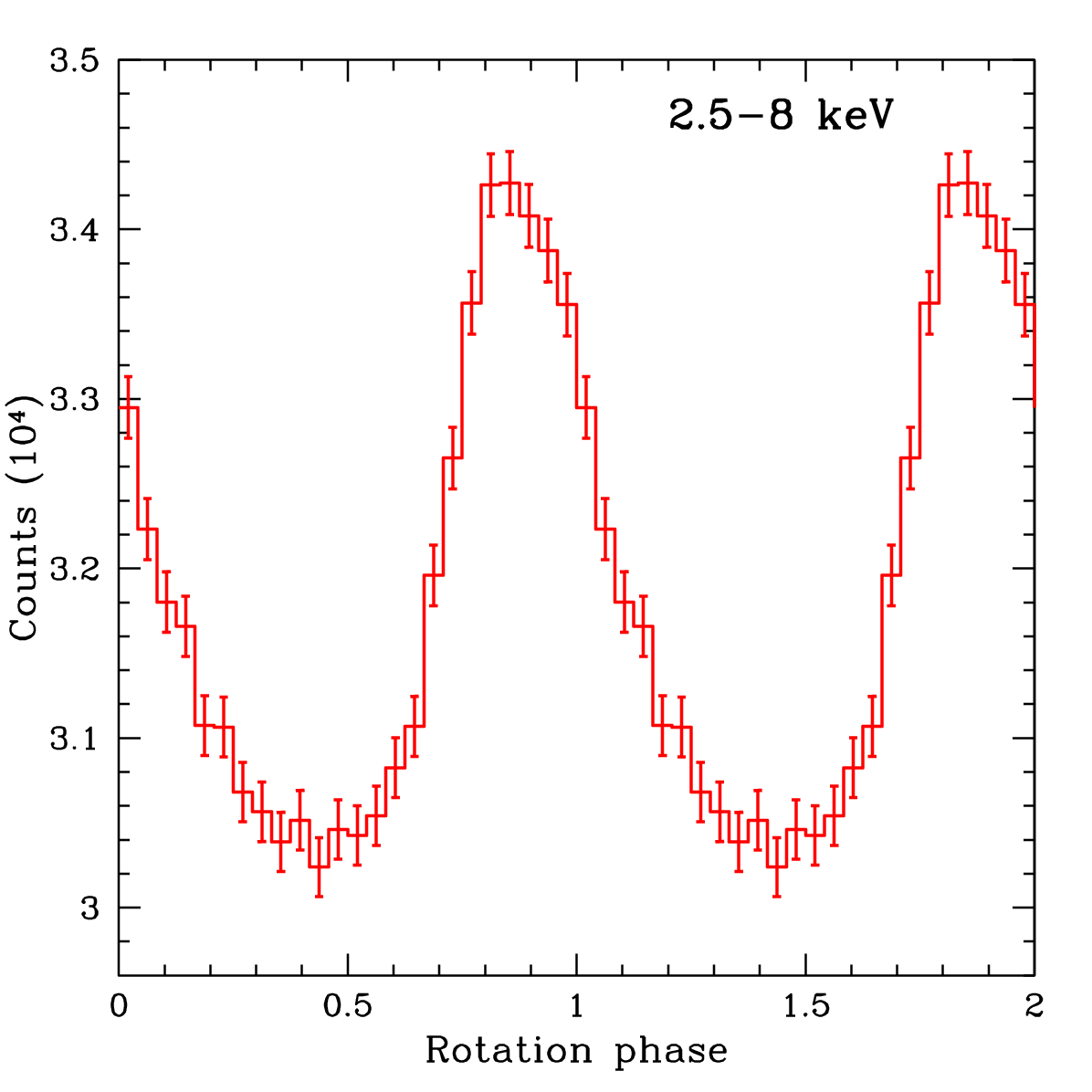}
\caption{Pulse profile of \psreightone\ from NICER data at 2.5-8~keV.
Errors are 1$\sigma$ uncertainty.
Two rotation cycles are shown for clarity.
\label{fig:1811_pp}}
\end{figure}

NICER monitoring of \psreightone\ began with a few observations in
2021 September and then commenced in earnest in 2022 March.
Here we report the timing model derived from all NICER data through
2025 May.
Analysis of these data suggest \psreightone\ suffers from timing noise.
As a result, we find the TOAs are better modelled by dividing them into
two epochs:
TOAs obtained before the period in late 2023 when \psreightone\ cannot
be observed due to its proximity to the Sun in the sky and
TOAs obtained after this period in early 2024,
i.e., before about MJD~60227 and after about MJD~60390.
Figure~\ref{fig:1811} shows residuals of our best-fit timing models
for these two epochs, which are given in Table~\ref{tab:1811}, and
Figure~\ref{fig:1811_pp} shows the pulse profile at 2.5--8~keV
from the combined NICER observations.
We note that we are able to construct a single timing model for the
entire period of NICER observations, but such a model requires two
additional spin terms (fourth and fifth time derivatives of $\nu$) and
yields a best-fit with RMS of 2.15~ms and $\chi^2/\mbox{dof}=90/35$.

\subsection{\psreightthree} \label{sec:psrj1813}

\begin{figure}
\includegraphics[width=\columnwidth]{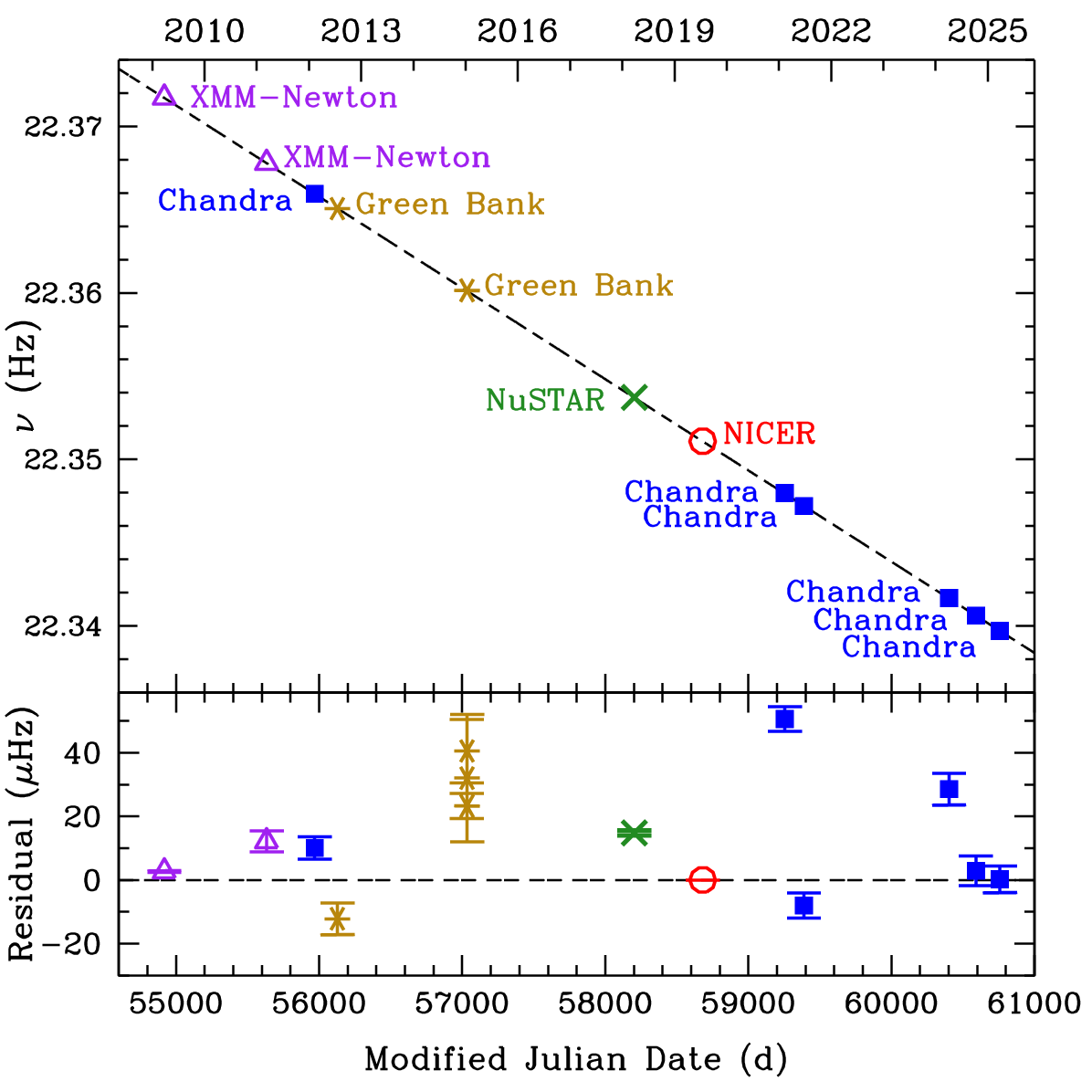}
\caption{Spin frequency $\nu$ of \psreightthree\ (top) and difference between
best-fit linear model and data (bottom) as functions of time.
Errors are 1$\sigma$ uncertainty.
Measurements of $\nu$ are made using XMM-Newton (triangles), Chandra (squares),
Green Bank Telescope in radio (stars), NuSTAR (cross), and NICER (circle).
Dashed line shows a linear fit of all $\nu$ measurements with
best-fit $\nudot=-6.3442\times10^{-11}\mbox{ Hz s$^{-1}$}$.
\label{fig:1813}}
\end{figure}

Unlike the rotation phase-connected (coherent) timing models for the
other pulsars presented in this work, the timing model for \psreightthree\
is incoherent, i.e., $\nudot$ is determined by a best-fit to a simple
linear model for the change of $\nu$ over time.  This is because of
the sparse number of individual measurements of the spin frequency such
that a phase-connected model is not possible.

\begin{table}
\centering
\caption{Incoherent timing parameters of \psreightthree.
Numbers in parentheses are 1$\sigma$ uncertainty in last digit.
The position is from VLA data (MJD 58119), with uncertainties of
$\sim0\fs009$ and $\sim0.13$~arcsec, and
proper motion has uncertainties of 3.7 and $6.7\mbox{ mas yr$^{-1}$}$ in
$\mu_\alpha\cos\delta$ and $\mu_\delta$, respectively \citep{dzibrodriguez21}.}
\label{tab:1813}
\begin{tabular}{lc}
\hline
Parameter & Value \\
\hline
R.A. $\alpha$ (J2000) & $18^{\rm h}13^{\rm m}35\fs173$ \\
Decl. $\delta$ (J2000) & $-17\degr49\arcmin57\farcs75$ \\
Solar system ephemeris & DE405 \\
Range of dates (MJD) & 54918.14$-$60758.00 \\
Epoch (MJD TDB) & 58681.04 \\
Frequency $\nu$ (Hz) & 22.3510838(2) \\
Freq.\ 1st derivative $\nudot$ (Hz s$^{-1}$) & $-6.3442(6)\times10^{-11}$ \\
Proper motion $\mu_\alpha\cos\delta$ (mas yr$^{-1}$) & $-5.0$ \\
Proper motion $\mu_\delta$ (mas yr$^{-1}$) & $-13.2$ \\
\hline
\end{tabular}
\end{table}

We report new spin frequency measurements of \psreightthree\ using three
recent Chandra CC observations taken in 2024 and 2025
(see Table~\ref{tab:data}).
Analysis of these data yield
$\nu=22.3416735(50)\mbox{ Hz}$ on MJD~60403.03,
and $\nu=22.3406187(47)\mbox{ Hz}$ on MJD~60590.78
and $\nu=22.3396994(42)\mbox{ Hz}$ on MJD~60758.00,
where the $1\sigma$ uncertainty in the last digit is given in parentheses.
These measurements are shown in Figure~\ref{fig:1813},
as well as previous ones
using Chandra and XMM-Newton \citep{halpernetal12,hoetal22},
Green Back Telescope at radio wavelengths \citep{camiloetal21},
NICER \citep{hoetal20a}, and NuSTAR \citep{hoetal22}.
Fitting a simple linear decline in spin frequency yields a best-fit
spin-down rate $\nudot=(-6.34424\pm0.00064)\times10^{-11}\mbox{ Hz s$^{-1}$}$.
Table~\ref{tab:1813} presents the resulting 16~yr timing model.
Residuals from the timing model,
as seen in the bottom panel of Figure~\ref{fig:1813},
suggest \psreightthree\ undergoes glitches with sizes as large as a
few tens of $\mu$Hz, which are typical for young pulsars.

We also attempt to measure the pulsed spectrum of \psreightthree\ by
combining all previous Chandra CC data.
This spectrum is obtained by first using measurements of the spin parameters
for each individual observation and extracting photons during the rotation
phases around the peak of the pulse profile (on-pulse)
and away from the peak (off-pulse).
Pulsed spectra are then determined by subtracting the off-pulse
emission from the on-pulse emission.
We then combine all the Chandra pulsed spectra into a single spectrum.
Unfortunately, the resulting
pulsed spectrum has poor photon statistics,
even when using a different extraction region,
and thus we do not show it nor fit it with spectral models.

\subsection{\psreight} \label{sec:psrj1849}

\begin{figure}
\includegraphics[width=\columnwidth]{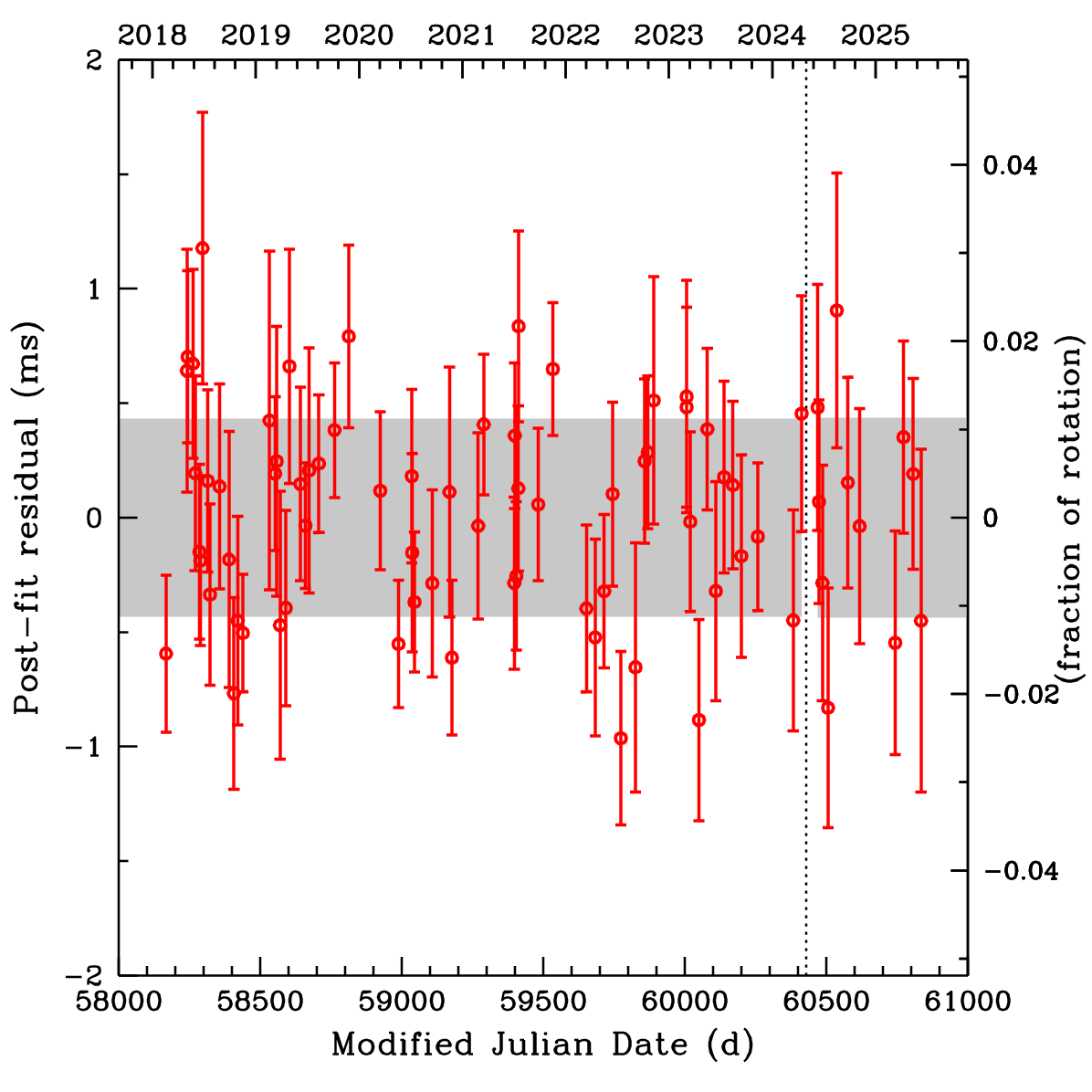}
\caption{Timing residuals of \psreight\ from two separate best-fits of
NICER pulse times-of-arrival,
topocentric TOAs before the glitch and barycentric TOAs after the glitch,
with the timing models given in the middle column of Table~\ref{tab:1849}.
Errors are 1$\sigma$ uncertainty.
The vertical dotted line indicates the approximate time of a spin-up glitch
(MJD~60428).
The RMS residual of each best-fit is illustrated by the shaded regions
between $\pm 0.432\mbox{ ms}$ before the glitch and
between $\pm 0.437\mbox{ ms}$ after the glitch.}
\label{fig:1849}
\end{figure}

\begin{table*}
\centering
\caption{Timing parameters of \psreight.
Numbers in parentheses are 1$\sigma$ uncertainty in last digit.
Position epoch is MJD~55885.
The middle column gives a timing model derived in two segments separated
by the glitch on MJD~60428; the first segment covers MJD~58166--60413,
includes position as a fit parameter, and uses topocentric TOAs, while
the second segment covers MJD~60469--60835 and uses barycentric TOAs.
The right column gives a timing model, including parameters of the glitch,
derived using only barycentric TOAs covering MJD~58923--60835.
}
\label{tab:1849}
\begin{tabular}{lcc}
\hline
Parameter & Value & Value \\
\hline
R.A. $\alpha$ (J2000) & $18^{\rm h}49^{\rm m}01\fs6194(26)$ & \\
Decl. $\delta$ (J2000) & $-00\degr01\arcmin17\farcs445(99)$ & \\
Solar system ephemeris & DE421 & \\
Range of dates (MJD) & 58166$-$60413 & \\
Epoch (MJD TDB) & 59289 & \\
$t_0$ (MJD) & 59289.3 & \\
Frequency $\nu$ (Hz) & 25.9581535113(2) & \\
Freq.\ 1st derivative $\nudot$ (Hz s$^{-1}$) & $-9.533059(9)\times10^{-12}$ & \\
Freq.\ 2nd derivative $\nuddot$ (Hz s$^{-2}$) & $5.12(9)\times10^{-23}$ & \\
Freq.\ 3rd derivative $\dddot{\nu}$ (Hz s$^{-3}$) & $-5.6(29)\times10^{-32}$ & \\
Freq.\ 4th derivative $\ddddot{\nu}$ (Hz s$^{-4}$) & $-1.03(34)\times10^{-38}$ \\
Freq.\ 5th derivative (Hz s$^{-5}$) & $-7.6(7)\times10^{-46}$ \\
Freq.\ 6th derivative (Hz s$^{-6}$) & $5.4(9)\times10^{-53}$ \\
RMS residual ($\mu$s) & 432 & \\
$\chi^2$/dof & 81.9/55 & \\
Number of TOAs & 65 & \\
\\
Glitch epoch (MJD) & & 60428$\pm$15 \\
$\Delta\phi$ & & $-0.16(96)$ \\
$\Delta\nu$ (Hz) & & $4.3841(42)\times10^{-5}$ \\
$\Delta\nudot$ (Hz s$^{-1}$) & & $-9.11(26)\times10^{-14}$ \\
$\Delta\nuddot$ (Hz s$^{-2}$) & & $6.61(89)\times10^{-22}$ \\
$\Delta\nu_{\rm d}$ (Hz) & & $5.5(52)\times10^{-7}$ \\
$\tau_{\rm d}$ (d) & & 30 \\
\\
Range of dates (MJD) & 60469$-$60835 & 58923$-$60835 \\
Epoch (MJD TDB) & 60651 & 60428 \\
$t_0$ (MJD) & 60617.3 & 59869.2 \\
Frequency $\nu$ (Hz) & 25.9570741758(13) & 25.9572155718(10) \\
Freq.\ 1st derivative $\nudot$ (Hz s$^{-1}$) & $-9.60605(53)\times10^{-12}$ & $-9.529343(68)\times10^{-12}$ \\
Freq.\ 2nd derivative $\nuddot$ (Hz s$^{-2}$) & $1.043(42)\times10^{-21}$ & $2.49(26)\times10^{-23}$ \\
Freq.\ 3rd derivative $\dddot{\nu}$ (Hz s$^{-3}$) & $-1.21(23)\times10^{-28}$ & $-2.54(41)\times10^{-30}$ \\
RMS residual ($\mu$s) & 437 & 508 \\
$\chi^2$/dof & 8.54/6 & 80.6/39 \\
Number of TOAs & 11 & 49 \\
\hline
\end{tabular}
\end{table*}

NICER observations of \psreight\ began on 2018 February 13, and
\citet{bogdanovetal19} reported a 1.5-year phase-connected timing
model using the first 7 months of NICER data as well as a
Swift observation from 2017 March.
\citet{hoetal22} extended the timing model to 4.7~yr
using NICER data through 2021 November.
\citet{kimetal24} further extended the timing model by about 8 months
using data through 2022 July.
Here we complete the timing model using the remaining effective NICER
data through 2025 June 8, covering a period of 7.3~yr that
\psreight\ was observed by NICER.

As done for \psrfour\ in Section~\ref{sec:psrj1412},
we seek to construct a timing model that includes the pulsar's position
as a fit parameter.  Thus we calculate 65 spacecraft topocentric TOAs 
from the dataset
before the glitch that occurred sometime between 2024 April 12 and May 12
(MJD~60412.6--60442.9) \citep{kuiperetal24}.
We fit these TOAs using PINT.
We do not include any proper motion of the pulsar since a recent analysis
comparing a Chandra 2012 ACIS-S image and a 2021 ACIS-I image yielded
a position change smaller than the uncertainties \citep{gagnonetal24}.
Meanwhile, since the post-glitch epoch covered by NICER data is short
and covers only 1~yr, we follow our more standard procedure and calculate
11 barycentric TOAs from the post-glitch data and fit these using TEMPO2
with a model that only includes frequency time derivatives through
$\nudddot$ and does not include position.

Figure~\ref{fig:1849} shows timing residuals of the TOAs used to obtain our
best-fit timing model, which is given in the middle column of Table~\ref{tab:1849}.
We find that our position from timing is consistent with the Chandra
HRC-S imaging position from \citet{kuiperhermsen15} of
R.A.~$=18^{\rm h}49^{\rm m}01\fs632$ and decl.~$=-00\degr01\arcmin17\farcs45$
at MJD 55885, which has a 90~percent confidence level uncertainty of 0.6~arcsec.
It is noteworthy that our timing position, with an uncertainty of 0.11~arcsec,
is now more accurate than the one from imaging.

To more accurately measure the glitch parameters, we perform a
second fit using barycentric TOAs and
restrict the timing model to terms up to $\nudddot$.
We are unable to obtain a single timing model covering the entire 7~years
of NICER data and therefore only use data since 2020 March 15 (MJD~58923).
The best-fit values are given in the right column of Table~\ref{tab:1849},
including those for the glitch.
The glitch magnitude, $\Delta\nu=43.8\mbox{ $\mu$Hz}$
and $\Delta\nu/\nu=1.69\times10^{-6}$, is among the largest observed
and comparable to those seen in the Vela pulsar and \psrfive\
(see, e.g., Figure~\ref{fig:0537glitch} and \citealt{fuentesetal17}).
We find the decay timescale $\tau_{\rm d}$
has an uncertainty of $\pm7\mbox{ d}$.

\begin{figure}
\includegraphics[width=\columnwidth]{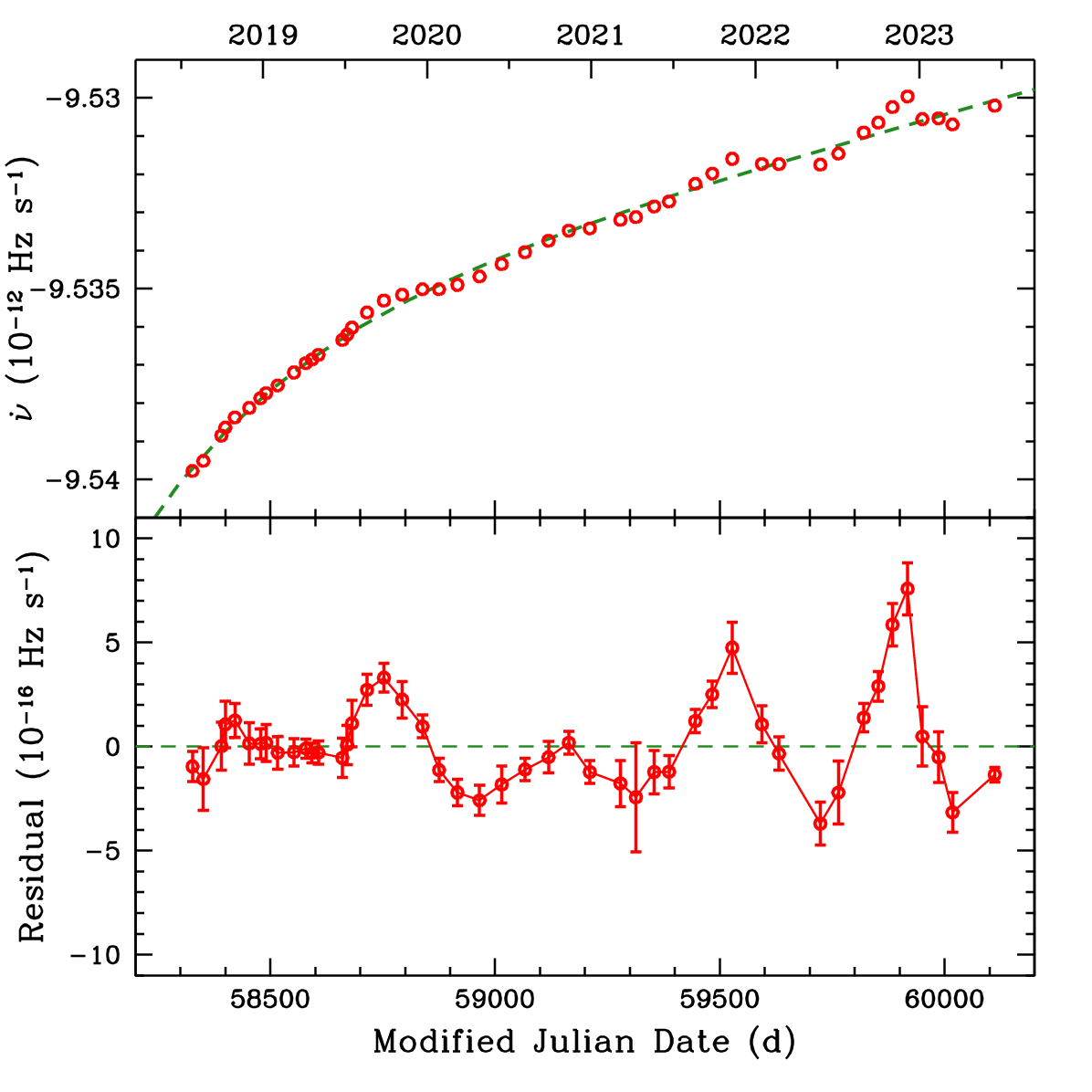}
\caption{Top panel shows spin-down rate $\nudot$ of \psreight\ calculated
from overlapping intervals at least 300~days long (see text for details).
The dashed line is for a model that includes $\nuddot$ and an exponential
decay with a $\sim 1\mbox{ yr}$ timescale.
Bottom panel shows difference between the data and model.
Errors are 1$\sigma$ uncertainty.
}
\label{fig:1849f1}
\end{figure}

Finally, Figure~\ref{fig:1849f1} shows the pre-glitch evolution of $\nudot$
and its residual relative to a model that includes an exponential decay,
although there does not appear to be evidence for a large glitch within
2000~d before the start of our NICER observations.
The spin-down values $\nudot$ shown here are calculated using at least
10~TOAs from overlapping intervals at least 300~days long with a varying
$\nuddot$, and each interval was moved by 5~days at each stride
(see \citealt{espinozaetal24}).
One can see the change of $\nudot$ over time and its oscillatory and
growing in amplitude deviation from the model with time.
This illustrates the long-term timing noise present in the spin evolution
of \psreight\ and the need for higher-order spin frequency time derivatives
in its timing model.

\section{Discussion} \label{sec:discuss}

We reported here the years-long timing models for seven young energetic
pulsars.  Six of the timing models are rotation phase-connected and make
use of all the data NICER collected on these pulsars during its 8~years
of operation.
For \psrzero, we detected its first three glitches in only 3.5 years
of observation.  It would be interesting to know if its glitching behavior
shows any similarities to that of the Big Glitcher, \psrfive.
For \psrfive, NICER data enabled detection of 23 glitches in 8 years,
comparable to the 45 glitches in 13 years found using RXTE.  These
glitches continue to show evidence for their predictability.
We also measured for the first time the spin pulsations and pulsed spectrum
of \psrfive\ using NuSTAR data.
For \psrzerofive, we provide its timing model over the past 10~years
using a combination of IXPE, NICER, and Swift data and see that its
braking index $n$ has been growing over this decade from $n\approx0$
to 1.6 as of mid-2025.
For \psrfour\ and \psreight, the more than 7~years of NICER monitoring
enabled measurements of their position from timing to accuracies comparable
to that obtained from imaging.

For all seven pulsars studied here, their timing models are crucial for
ongoing efforts to detect gravitational waves emitted by pulsars.
These gravitational waves are continuous signals that can be emitted
for the entire lifetime of a pulsar
(see \citealt{haskellbejger23,riles23,wette23}, for reviews).
The frequency range of current ground-based gravitational wave detectors
means that they are only sensitive to fast-spinning pulsars, and
the most sensitive searches are those targeting pulsars that have an
accurate timing model like those derived here.
In fact, the timing models for the seven pulsars presented here and in
previous works have already been used to search for gravitational waves
in LIGO/Virgo/KAGRA data
\citep{abbottetal21a,abbottetal21b,abbottetal22a,abbottetal22b,abacetal25}.

Finally, it is important to emphasize that
five of our seven pulsars are only observed as pulsars in the X-ray and
thus their timing models necessitate regular monitoring by sensitive X-ray
telescopes with fast and accurate timing capabilities, such as NICER.
As an example, we searched for but did not detect pulsed emission
from \psreightone\ using Fermi-LAT data, like the non-detections of
\psrfour\ and \psreight\ described in \citet{hoetal22};
we also reported in that work a possible detection of \psrzero\ in Fermi-LAT
data, but our much longer time baseline now no longer supports detection.
Potential new general purpose X-ray telescopes,
such as AXIS \citep{reynoldsetal24},
could be invaluable in this regard if they possess an observing mode
that utilizes a subarray and/or fast readout.
The capability to obtain timing models for X-ray pulsars is required
for the most sensitive searches for gravitational waves using the next
generation gravitational wave detectors being developed for the 2030s,
such as Cosmic Explorer \citep{evansetal21} and
Einstein Telescope \citep{maggioreetal20}.

\section*{Acknowledgements}

WCGH thanks Paul Ray for discussions on timing analyses and
George Younes for discussions on NuSTAR data analysis.
WCGH acknowledges support through grants 80NSSC22K1305, 80NSSC24K1195,
and 80NSSC25K0085 from NASA and Chandra award SAO GO4-25038X.
Chandra grants are issued by the Chandra X-ray Center (CXC),
which is operated by the Smithsonian Astrophysical Observatory
for and on behalf of NASA under contract NAS8-03060.
CME acknowledges support from ANID/FONDECYT, grant 1211964.
TL acknowledges financial support from the Haverford Koshland
Integrated Natural Sciences Center (KINSC).
BW acknowledges financial support from the Boughn-Gollub-Partridge
Fund at Haverford College.
SG acknowledges the support of the CNES.

This research made use of data obtained from the Chandra Data
Archive and the Chandra Source Catalog, and software provided by the
Chandra X-ray Center (CXC) in the application packages CIAO and Sherpa.
This work is supported by NASA through the NICER mission and the
Astrophysics Explorers Program and uses data and software provided
by the High Energy Astrophysics Science Archive Research Center
(HEASARC), which is a service of the Astrophysics Science Division
at NASA/GSFC and High Energy Astrophysics Division of the
Smithsonian Astrophysical Observatory.
This research made use of data from the NuSTAR mission, a project led by
the California Institute of Technology, managed by the Jet Propulsion
Laboratory, and funded by the NASA. Data analysis was performed using
the NuSTAR Data Analysis Software (NuSTARDAS), jointly developed by the
ASI Science Data Center (SSDC, Italy) and the California Institute of
Technology (USA).
This work made extensive use of the NASA Astrophysics Data System
(ADS) Bibliographic Services and the arXiv.

\section*{Data Availability}

Data underlying this article will be shared on reasonable request
to the corresponding author.
 

\bibliographystyle{mnras}
\bibliography{arxiv}






\bsp
\label{lastpage}
\end{document}